\newif\ifsingle
\ifsingle
\documentclass[11pt,draftclsnofoot, onecolumn]{IEEEtran}		
\else		
\documentclass[10pt,final, twocolumn]{IEEEtran}
\fi
 
\usepackage{amsmath,graphicx}
\usepackage{amssymb}
\usepackage{acronym}

\usepackage[belowskip=-10pt,aboveskip=1.8pt]{caption}
\usepackage{standalone}
\usepackage{subcaption}
\usepackage{tikz}
\usepackage{url,cite}
\usepackage{verbatim, cases}
\usepackage[bookmarks,colorlinks]{hyperref}
\usepackage{soul, xcolor, flushend}
\usepackage{soul}
\usepackage{placeins}
\usepackage{enumitem}

\usepackage[ruled,linesnumbered,vlined]{algorithm2e}
\SetKwInput{KwData}{\textbf{Init}} 

\newcommand{\myVec}[1]{{\boldsymbol{#1}}}

\newcommand{\mySet}[1]{\mathcal{#1}}
\newcommand{\PathLoss}{\Gamma}

\newtheorem{lemma}{Lemma}

\usepackage[all=normal,paragraphs=tight,floats=normal,mathspacing=normal,wordspacing=tight,charwidths=tight,mathdisplays=normal,leading=normal]{savetrees}
\definecolor{NewColor}{rgb}{0,0,0}

\acrodef{lwa}[LWA]{leaky wave antenna}
\acrodef{los}[LOS]{line-of-sight}
\acrodef{hmm}[HMM]{hidden Markov model}
\acrodef{dl}[DL]{deep learning}
\acrodef{dnn}[DNN]{deep neural network}
\acrodef{snr}[SNR]{signal-to-noise ratio}
\acrodef{bs}[BS]{base station} 
\acrodef{cpu}[CPU]{centralized processing unit} 
\acrodef{mimo}[MIMO]{multiple-input multiple-output}
\acrodef{awgn}[AWGN]{additive white Gaussian noise} 
\acrodef{cpu}[CPU]{central processing unit} 
\acrodef{ml}[ML]{machine learning} 
\acrodef{mse}[MSE]{mean-squared error}
\acrodef{iot}[IOT]{Internet of Things}
\acrodef{rmse}[RMSE]{root mean squared error}
\acrodef{rmspe}[RMSPE]{root mean squared periodic error}
\acrodef{mmse}[MMSE]{{minimum mean-squared error}}
\acrodef{lmmse}[LMMSE]{{linear} MMSE}
\acrodef{mle}[MLE]{maximum likelihood estimation}
\acrodef{admm}[ADMM]{alternating direction method of multipliers}
\acrodef{dadmm}[D-ADMM]{distributed alternating direction method of multipliers}
\acrodef{sar}[SAR]{successive approximation register}
\acrodef{adc}[ADC]{analog-to-digital converter} 
\acrodef{dac}[DAC]{digital-to-analog converter} 
\acrodef{msb}[MSB]{most significant bit}
\acrodef{lsb}[LSB]{least significant bit}
\acrodef{sh}[S\&H]{sample-and-hold}
\acrodef{aod}[AoD]{angle of departure}
\acrodefplural{aod}[AoDs]{angles of departure}
\acrodef{ofdm}[OFDM]{orthogonal frequency-division multiplexing}
\acrodef{ofdma}[OFDMA]{orthogonal frequency-division multiple access}
\acrodef{ga}[GA]{genetic algorithm}

\acrodef{aoa}[AoA]{Angle of Arrival}
\acrodefplural{aoa}[AoAs]{Angles of Arrival}
\acrodef{em}[EM]{electromagnetic}
\acrodef{cmos}[CMOS]{complementary metal-oxide semiconductor}
\acrodef{ttd}[TTD]{ true-time-delay}

\usetikzlibrary{trees,arrows}

\setstcolor{blue}

\title{Wideband THz  Multi-User Downlink Communications with Leaky Wave Antennas
}

\author{
\IEEEauthorblockN{Natalie Lang, Yaela Gabay, Nir Shlezinger, \IEEEmembership{Senior Member, IEEE}, Tirza Routtenberg, \IEEEmembership{Senior Member, IEEE}, Yasaman Ghasempour,  George C. Alexandropoulos, \IEEEmembership{Senior Member, IEEE}, and Yonina C. Eldar, \IEEEmembership{Fellow, IEEE} 
\thanks{
Parts of this work were presented at the IEEE International Conference on Acoustics, Speech, and Signal Processing (ICASSP) 2024 as the paper \cite{gabay2023leaky}.
N. Lang, Y. Gabay, N. Shlezinger, and T. Routtenberg with the School of ECE, Ben-Gurion University of the Negev, Israel (e-mail: \{langn; yaelag\}@post.bgu.ac.il; \{nirshl; tirzar\}@bgu.ac.il). 
Y. Ghasempour is with the Department of ECE, Princeton University, NJ, USA (e-mail: ghasempour@princeton.edu). 
	G. C. Alexandropoulos is with the Department of Informatics and Telecommunications, National and Kapodistrian University of Athens, 15784 Athens, Greece and the Department of Electrical and Computer Engineering, University of Illinois Chicago, Chicago, IL 60601, USA (e-mail: alexandg@di.uoa.gr). 
    Y. C. Eldar is with the Faculty of Math and CS, Weizmann Institute of Science, Rehovot, Israel (e-mail: yonina@weizmann.ac.il).
\\
This work was partially supported by the Israeli Ministry of National Infrastructure, Energy, and Water Resources, the SNS JU project TERRAMETA under the European Union's Horizon Europe research, innovation programme under Grant Agreement No 101097101, and the US National Science Foundation under Grant Number CNS-2145240.}}}

\begin{document}
%\ninept

\maketitle
%
%%%%%%%%%%%%%%%%
%%% Abstract %%%
%%%%%%%%%%%%%%%%
%
\begin{abstract}
Future wireless systems are envisioned to utilize the large spectra available at THz bands for wireless communications. Extremely massive multiple-input multiple-output (MIMO) antennas can be costly and power inefficient for wideband THz communications. An alternative antenna technology, which can achieve low-cost and power-efficient THz signaling, is based on \acp{lwa}. In this paper, we explore the usage of the \acp{lwa} for wideband downlink multi-user THz communications. We propose a model for \ac{lwa}-aided communication systems that faithfully captures the antenna operations. We \textcolor{NewColor}{show} that \acp{lwa} yield frequency-dependent beams, where the equivalent wideband channel induces a dependence between angle, frequency, and spectral lobe width. We identify the \ac{lwa}’s inherent frequency-selective beamsteering capabilities as motivating multi-band THz communications, 
\textcolor{NewColor}{in which subbands are allocated among users based on their relative angles.} Then, we propose an alternating optimization algorithm for jointly optimizing the \ac{lwa} configuration along with the spectral division and power allocation to maximize the achievable sum rate performance. Our numerical results show that a single \ac{lwa} can generate diverse beampatterns, exhibiting performance comparable to costly \textcolor{NewColor}{MIMO architectures in wideband THz multi-user systems}. 
\end{abstract}

\acresetall

%----------------------------------------------------------------------------------------
%	INTRODUCTION
%----------------------------------------------------------------------------------------

\section{Introduction}
\label{sec:intro}

Wireless communications are subject to constantly growing demands in throughput, connectivity, and latency~\cite{giordani2020toward}. To meet these requirements, future wireless systems will utilize high-frequency regimes, and particularly explore the abundant bandwidth available at THz bands~\cite{wang2021key,jiang2024terahertz}. Indeed, the exploration of such wide bandwidth is expected to overcome the spectral congestion of legacy bands~\cite{saad2019vision,rappaport2019wireless}, and support the throughput requirements of the upcoming sixth generation (6G) wireless technologies~\cite{rajatheva2020white}.

Despite the promising spectral resources that can be provided by THz bands, designing communication systems that operate in such high frequencies induces critical design challenges in terms of hardware, power, and energy efficiency. Consequently, THz systems are likely to deviate from conventional fully digital \ac{mimo} antenna architectures employed in, e.g., {legacy} sub-6 GHz bands~\cite{polese2020toward}. Such traditional architectures, comprised of multiple antenna elements, each connected with a dedicated RF chain, are expected to be extremely costly in THz bands~\cite{akyildiz2022terahertz}.  
\textcolor{NewColor}{ Recent studies have investigated the interplay between antenna design and propagation in the THz band, emphasizing the importance of antenna-specific channel models and their implications for next generation networks \cite{Sen2024,Guven2024}. Additionally, novel LWA designs offering wideband radiation have been explored, further highlighting the potential of LWA technology for high-frequency applications \cite{Comite2018,monnai2023terahertz}.}

To date, several different architectures for THz systems are considered in the literature. A common approach assumes hybrid analog/digital \ac{mimo} transceivers~\cite{thanh2023deep,berman2021resource, ghasempour2018decoupling,RIS_THz_Eucap,TERRAMETA_switches}. However, these architectures typically rely on phase shifters~\cite{ioushua2019family}, which can be limiting factors in terms of power consumption and frequency dependence in THz bands. Additionally, for a multi-user multi-carrier THz communication system, there are strict limitations over the number of directed beams that can be formed simultaneously by a given number of antennas in traditional arrays. Thus, the
\textcolor{NewColor}{number of beams that can be generated with} a specific fixed antenna configuration is limited.
A related family of approaches implements antennas using metasurfaces~\cite{shlezinger2021dynamic,shao2023hybrid, chen2022m,DMA_EE,DMA_NF_tracking}, possibly as a form of holographic \ac{mimo}~\cite{huang2020holographic}. However,  such architectures are typically not designed for wideband THz signaling and have been primarily conceptually studied for the narrowband case \cite{HMIMO_survey}.

Another candidate architecture is based on \ac{ttd} circuitry in hybrid \ac{mimo}, either replacing phase shifters~\cite{ghaderi2019integrated} or in addition to them~\cite{zhai2020thzprism,zhai2021ss,dai2022delay,nguyen2024joint,do2024hybrid}. Such architectures can create several on-demand frequency-dependent beams~\cite{li2022rainbow,ratnam2022joint} to serve users at a wide range of angles. These spatially spectral beams are fundamentally different from small dispersion effects (known as beam squint) in traditional phased arrays, which are usually regarded as unwanted side effects causing severe array gain loss~\cite{zhai2020thzprism,zhai2021ss,dai2022delay,nguyen2024joint,do2024hybrid,THz_CE_beamsquint}. In~\cite{dai2022delay}, a \ac{ttd}-based precoder was presented, realizing improved achievable rates and energy efficiency than hybrid precoding, and ~\cite{li2022rainbow,ratnam2022joint} proposed a \ac{ttd}-based system architecture that was matched with an access control protocol, synchronizing multi-user communications under a fixed beam configuration. While these works indicate the potential gains of exploiting frequency-selective beams in THz regime, the increased cost, hardware complexity, and power consumption associated with numerous phase-shifters and \acp{ttd} circuitry required for these architectures hinder their adoption in practical settings.  

An alternative emerging antenna technology that is particularly suitable for THz \textcolor{NewColor}{signaling} is based on leaky waveguides~\cite{oliner2007leaky, guerboukha2023conformal, zetterstrom2019low}. \Acp{lwa} enable frequency-selective beamsteering of extremely wideband signals, creating the so-called THz rainbow~\cite{ghasempour2020single}. The beampatterns generated by \Acp{lwa} resemble those produced by hybrid \ac{mimo} transmitters utilizing \ac{ttd} and phase shifters~\cite{zhai2020thzprism,zhai2021ss, dai2022delay,nguyen2024joint,do2024hybrid}, albeit in a cost and power efficient manner~\cite{kludze2023leakyscatter,karl2015frequency,rahmani2023next}. Despite their potential for THz communication systems, \acp{lwa} are mostly studied in terms of localization~\cite{ghasempour2020leakytrack,kludze2022quasi}, \textcolor{NewColor}{retrodirective backscattering~\cite{kludze2024frequency}, and} wireless security~\cite{yeh2023security}, while their transmission model and ability to support wideband multi-user wireless communications have not yet been studied.

In this work, we consider \ac{lwa}-aided THz communications focusing on multi-user downlink systems. We propose a transmission model for \ac{lwa} signaling that enables exploring their beamforming capabilities for multi-user communications, while faithfully capturing the antenna operations, as physically formulated and experimentally validated in ~\cite{ghasempour2020single}.
We identify two design parameters of \acp{lwa}: the plate separation and slit length, which directly impact the three-dimensional (3D) angle-frequency relationship, and the resulting frequency-dependent directional beams. Looking into the resulting beampatterns, we identify an inherent lack of symmetry in the induced channel's behavior towards frequency changes, which motivates us to 
\textcolor{NewColor}{optimize the subbands division for utilizing \ac{ofdma} \cite[Ch. 10]{Hossain_Rasti_Le_2017} based on the relative angles of the users}. We then propose an algorithm based on alternating optimization for jointly tuning the \ac{lwa} parameters alongside the digital power allocation, and the spectral division for wideband signaling; with the goal of maximizing the achievable sum rate under physical constraints. We numerically show that a single \ac{lwa} can form various beampatterns that support multiple downlink users, achieving sum rate performance that is comparable with fully digital \ac{mimo} systems that are equipped with multiple antennas and complex power-demanding phase shifting/time delay capabilities~\cite{boljanovic2020true}. %\textcolor{red}{much less power  [Yasaman: can we quantify this? back this with reference?]}. 

Our main contributions are summarized as follows:
\begin{itemize}
    \item \textbf{\ac{lwa}-aided communication model:}
    We formulate a physically compliant model for downlink wideband THz communications, where transmission is carried out using an \ac{lwa}. Our model describes physical characterizations of leaky \textcolor{NewColor}{wave antennas}
    via a mathematical model that supports studies from a communications theory perspective.
    \item \textbf{Transmission and \ac{lwa} optimization:} We propose an algorithm for jointly optimizing the internal \ac{lwa} parameters along with the \textcolor{NewColor}{power division, and} the spectral allocation. \textcolor{NewColor}{The latter is geared towards \ac{ofdma} systems, and is inspired by our modeling of \ac{lwa}-based channels which unveils that there is inherent coupling between the relative angle and the attenuated subbands.} This algorithm employs alternating optimization on the design parameters, with dedicated tools to tune each parameter, and is accompanied by a complexity analysis.
    \item \textbf{Extensive experimentation:} We extensively evaluate the potential of \acp{lwa} in facilitating cost-efficient and sustainable THz wireless communications. Our numerical studies reveal that a properly tuned \ac{lwa} can achieve multiple directed beams and approach the rates supported by costly conventional fully digital \ac{mimo} transmitters. \textcolor{NewColor}{We also show that our \ac{ofdma}-oriented subband division allows achieving performance within a minor gap from that achieved when having all users share the complete spectra, which requires sophisticated broadcast channel coding.} 
\end{itemize}

The rest of the paper is organized as follows: Section~\ref{sec:System Model} presents the proposed \ac{lwa} communication model. Joint tuning of the \ac{lwa} and power allocation is presented and evaluated in Sections~\ref{sec:Beamforming} and \ref{sec:Experimental Study}, respectively, while Section~\ref{sec:conclusions} concludes the paper.

%----------------------------------------------------------------------------------------
%	System Model
%----------------------------------------------------------------------------------------

\section{System Model and Problem Formulation}
\label{sec:System Model}
In this section, we describe the system model underlying our study. We begin by detailing the characteristics and operational principles of  \acp{lwa} in Subsection \ref{ssec:LWA}. Then, we present the resulting \ac{lwa}-aided THz communication system in Subsection~\ref{ssec:dowlink}, based on which we formulate the problem of optimizing the \ac{lwa} parameters for enhanced downlink communications in Subsection \ref{ssec:Problem}.

%%%%%%%%%%%%%%%%%%%%%%%%%%%%%%%%%%%
%%%	Leaky Waveguide Antenna Model
%%%%%%%%%%%%%%%%%%%%%%%%%%%%%%%%%%%
\subsection{LWA Modeling}
\label{ssec:LWA} 
\begin{figure}
		\centering
		\includegraphics[width=0.80\linewidth]{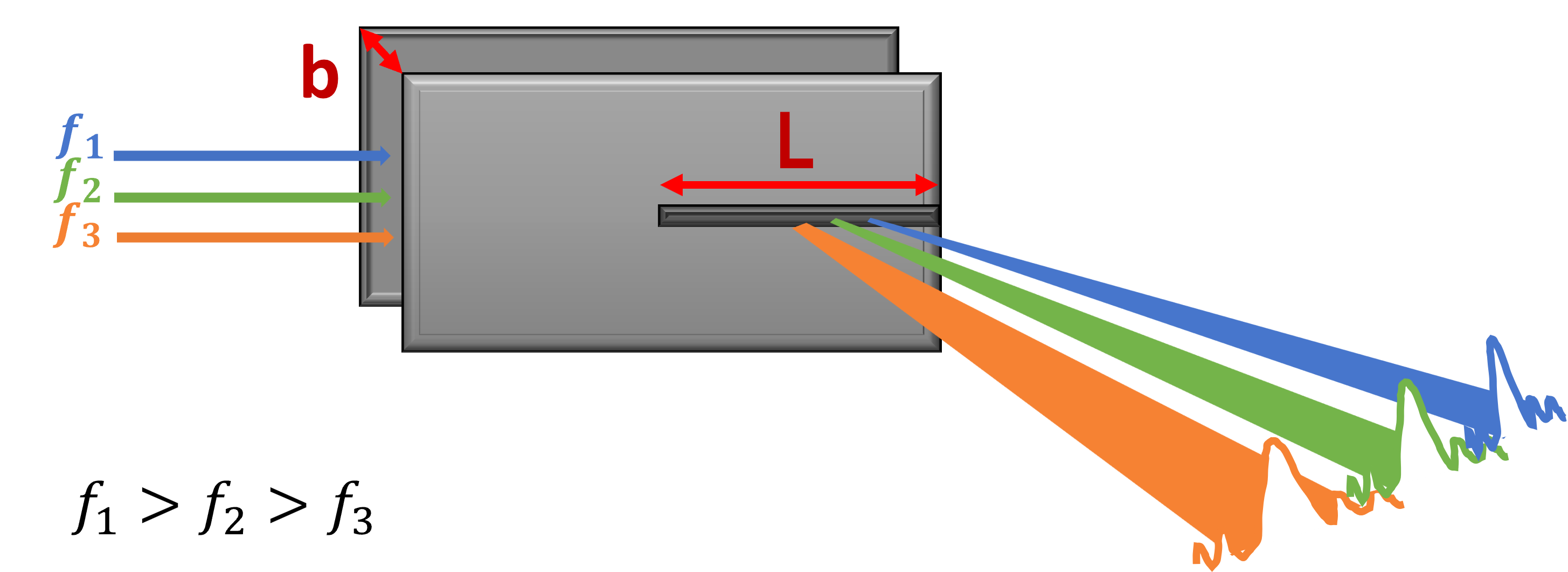}
		\caption{Beamforming of different frequency components using an \ac{lwa} with plate separation $b$ and slit length $L$.} 
  \label{fig:LWA}
\end{figure}
\textcolor{NewColor}{At THz frequencies, \acp{lwa} are commonly constructed using}
%In its most general form, an \ac{lwa} consists of 
two parallel metal plates,
\textcolor{NewColor}{although other implementations exist, including those}
%(though it can also be
fabricated using typical \ac{cmos} technology \cite{CMOS_LWA}.
\textcolor{NewColor}{Parallel-plate waveguides have been experimentally demonstrated for \ac{lwa} applications, with studies highlighting their simplicity and low-cost fabrication~\cite{ghasempour2020single, ghasempour2020leakytrack, guerboukha2023conformal}.}
The plates are separated by a distance denoted by $b$, and one of the plates has a slit of length $L$, as illustrated in Fig.~\ref{fig:LWA}. 
\textcolor{NewColor}{While this structure is relatively bulky, it is inexpensive to produce, and similar designs have been integrated with CMOS-based transceivers~\cite{THz_Prism_2021}.}
As a wideband THz signal travels between the plates, it emerges from the slit at various \acp{aod} \cite{ghasempour2020single}. The emission angle changes monotonically for different frequency components, creating a {\em THz rainbow} effect at the \ac{lwa} output. The propagation profile of the signal along the $L$-length slit radiates spectral components in directed beams in a manner that bears some similarity to synthetic apertures generated by moving devices~\cite{moreira2013tutorial}. To formulate this mathematically, a spectral component at frequency $f$ is radiated at azimuthal angle $\phi_f$, as follows~\cite{ghasempour2020single}:
\begin{equation} 
\label{eq:frequency-angle relation}
\phi_f =\sin^{-1}\left( \frac{ c}{ 2bf}\right),   
\end{equation}
where $c$ is the speed of light. 
Since $\phi_f$ is inversely proportional to $f$, it can be easily concluded that {\em{higher}} frequencies are emitted at {\em{smaller}} angles relative to the plate's axis.

The angular width of each directed monochromatic beam around its maximum emission angle depends on the \ac{lwa} parameters $L$ and $b$, and \textcolor{NewColor}{decreases} as the frequency $f$ increases.  
In particular, \textcolor{NewColor}{as experimentally shown in \cite{ghasempour2020single}}, the diffraction pattern at angle $\phi$ and frequency $f$ can be faithfully modeled as \cite{kludze2022quasi,sutinjo2008radiation}: 
\begin{subequations}
\label{eq:angular width}
    \begin{align}
    &G(\phi,f ; b,L) =\textcolor{NewColor}{L} {\rm sinc}\left[\left(\beta-j\alpha-k_0\cos\phi\right)\frac{L}{2}\right],
    \label{eq:angular widthA}\\
     &\beta \triangleq k_0\sqrt{1-\left(\frac{c}{2bf}\right)^2}.
     \end{align}
\end{subequations}
In \eqref{eq:angular width},  $k_0\triangleq\frac{2\pi f}{c}$ is the free-space wave number, and $\alpha$ is a parameter describing the energy loss of the transmitted wave as a result of leakage out of the LWA's slit. \textcolor{NewColor}{As shown in \cite{ghasempour2020single},} this loss is typically negligible, i.e., $\alpha \ll \beta$, and thus, in the following we use $\alpha=0$.  

It follows from the diffraction pattern in \eqref{eq:angular width} that a single \ac{lwa} can steer THz beams by tuning its $L$ and $b$ physical parameters. Moreover, signal components at different frequencies are radiated at different angles and with different-width beampatterns. This indicates that by optimizing the tuning of the \ac{lwa} parameters as well as the spectral allocation of the transmitted signal, this antenna structure can efficiently support high-frequency communications. In particular, as we will study in the sequel, \acp{lwa} are capable of handling multi-user communication scenarios with diverse beam patterns and high data rates, while using significantly less power compared to fully digital \ac{mimo} systems.   

%%%%%%%%%%%%%%%%%%%%%%%%%%%
%%%	{LWA-Aided Downlink Communications
%%%%%%%%%%%%%%%%%%%%%%%%%%%
\subsection{LWA-Aided Downlink Communications}
\label{ssec:dowlink}  
\begin{figure}
		\centering
		\includegraphics[width=1\linewidth]{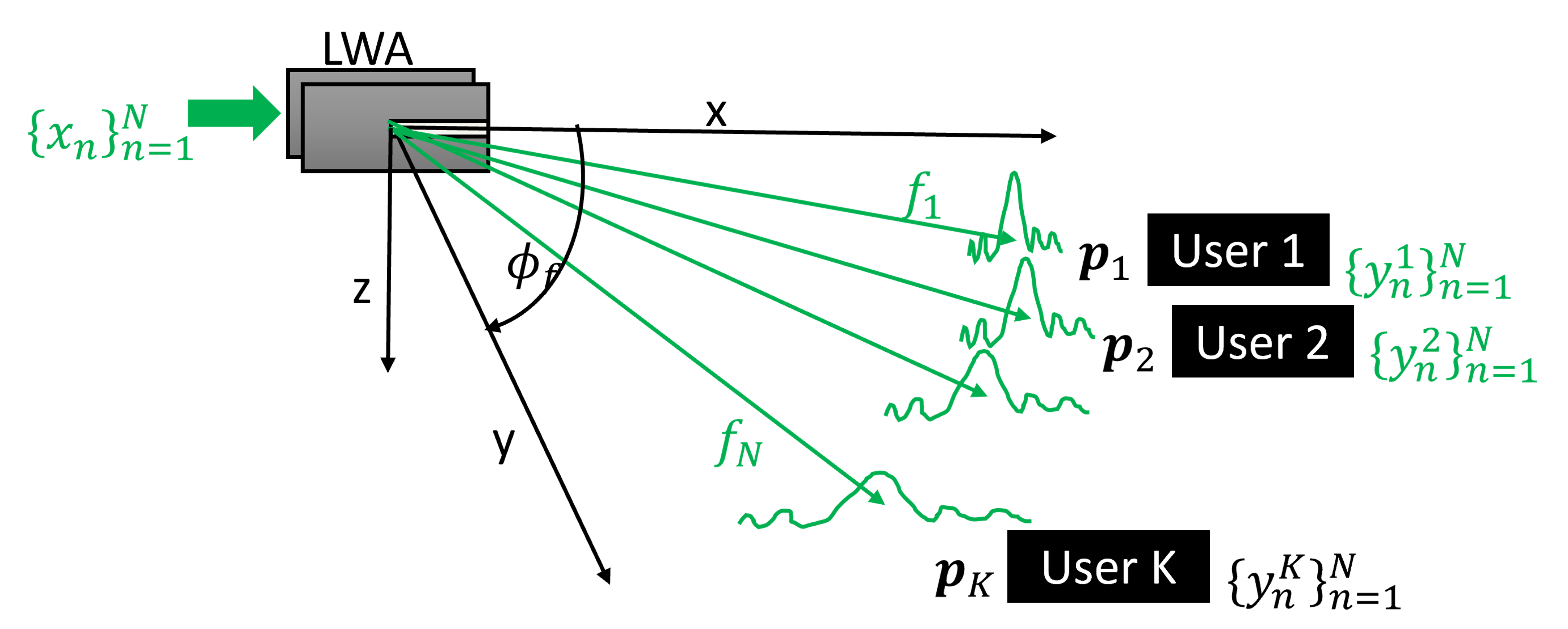}
		\caption{Downlink THz communications with a single \ac{lwa} transmitting to $K$ users using $N$ frequency bins.} 
            \label{fig:Channel}
\end{figure}
We consider a multi-user downlink setup where a \ac{bs}, equipped with a single \ac{lwa}, wishes to serve $K$ single-antenna users. We study only azimuthal steering in our \ac{lwa} modeling, assuming all users are located at the same height. We focus on \ac{los} settings, which is the expected operating regime of THz systems \cite{han2022terahertz}, where each $k$th user ($k=1,\ldots,K$) is located at a relative distance $\rho_k$ and angle $\varphi_k$ from the \ac{bs}. The \ac{bs} employs wideband signaling with $N$ non-overlapping subbands with central frequencies $f_1,\ldots,f_N$ \textcolor{NewColor}{that uniformly divide the operational band  $BW$, i.e., the subarrier spacing is $BW/N$.}
% . We denote the subband related to frequency $f_n$ by $\Delta f_n$ and assume that:
% \begin{equation}
% \label{eq:subband constraint}
%     \sum _{n=1}^N \Delta f_n = BW = f_{\max}-f_{\min},
% \end{equation}
% where $BW$ represents the operational bandwidth, whose lower and upper boundaries are the frequencies $f_{\min}$ and $f_{\max}$, respectively. 
The signal modulated at each $n$th subband ($n=1,\ldots,N$) is represented by $x_n\in \mathbb{C}$ (in practice chosen from a discrete modulation set), and its power is denoted by $ P_n \geq 0$. In addition, the \ac{bs} is subject to a total power constraint $P$, such that it holds that:
\begin{equation}
\label{eq:Power const.}
\sum _{n=1}^N P_{n} \leq P.
\end{equation}

The use of an \ac{lwa} implies that the spectral component at each frequency $ f_{n} $ is emitted at an angle $ \phi _{f _{n}}$, as depicted in  Fig.~\ref{fig:Channel}. The \ac{lwa} induces a channel whose coherence bandwidth changes inherently with frequency/angle in a manner where {higher} frequencies or low angles are translated into {narrower} beams, which in turn impact the coherence bandwidth.  
For simplicity, we assume that each subband is not larger than its corresponding coherence bandwidth, namely, that $BW/N$ is sufficiently narrow such that \eqref{eq:frequency-angle relation} and \eqref{eq:angular width} hold for the entire subband centered at $f_n$ \cite[Ch. 2]{tse2005fundamentals}.
% To this end, by denoting with $d_n$ the coherence bandwidth around $f_n$, we impose the following constraint on the subband widths:
% \begin{equation}
% \label{eq:subband upper constraint.}
% 0 \leq \Delta f_n \leq d_{n},   \quad\forall n\in \{1, \ldots, N\}.
% \end{equation}

Let $y_n^k \in \mathbb{C}$ represent the signal received by each $k$th user at each $n$th subband. Accordingly, we introduce the channel input-output relationship, which, based on the \ac{los} assumption, we model as follows:
\begin{equation} 
\label{eq:channel input-output}
{y}_{n}^k=   G(\varphi _{k},f_{n} ; b,L) \PathLoss(\rho _{k}, f _{n}) x_{n}+{w}_{n}^k, \quad n=1,\ldots,N,
\end{equation}
where $\PathLoss(\cdot,\cdot)$ is an attenuation coefficient that depends on relative distance $\rho_k$ and the frequency $f _{n}$, and ${w}_{n}^k$ is spectrally and temporally independent and identically distributed Gaussian noise. 
\textcolor{NewColor}{The noise is white with power spectral density $\sigma ^{2}$, thus, the variance of ${w}_{n}^k$  is  $\frac{BW}{N}\sigma ^{2}$.}
 The function $G(\varphi_k,f_n ; b,L)$ is the \ac{lwa} diffraction pattern at angle $\varphi_k$ and frequency $f_n$, defined in \eqref{eq:angular width}, and is the key component relating the resulting communications channel (defined in Subsection~\ref{ssec:Problem}) to the \ac{lwa} parameters $b,L$.

\color{NewColor}{We consider two approaches to support multi-user communications over the available bandwidth:
\begin{enumerate}[label=\textbf{A\arabic*)}, wide=0pt, leftmargin=*]  
    \item \label{itm:OFDM} The first approach has the \ac{bs} employ \ac{ofdm} multi-band signaling, while allowing all the users to simultaneously utilize the entire frequency band. While this form of non-orthogonal multi-user operation is typically preferable from a network information theoretic perspective~\cite{el2011network}, it entails excessive complexity in terms of coding and receiver operation. 
    \item \label{itm:OFDMA} Accordingly, the second approach divides the subbands between the users via \ac{ofdma} \cite[Ch. 10]{Hossain_Rasti_Le_2017}. Here, each subband is utilized only by a single user, allowing the usage of simple coding schemes utilized in standard orthogonal systems.  
\end{enumerate}
To encompass both approaches, we use $\mySet{S}_n\subseteq \mySet{K}\triangleq \{1,\ldots,K\}$ to denote the set of users that utilize the $n$th subband. Now, \ref{itm:OFDM} and \ref{itm:OFDMA} can be distinguished by the cardinality of $\mySet{S}_n$, as
\begin{equation}\label{eq: OFDMA_setting}
    |\mySet{S}_n| = \begin{cases}
        K & \ref{itm:OFDM} \\
        1 & \ref{itm:OFDMA}
    \end{cases}, \quad \forall n\in \{1,\ldots,N\}.
\end{equation}
\color{black}

%%%%%%%%%%%%%%%%%%%%%%%%%%%
%%%	Problem Formulation
%%%%%%%%%%%%%%%%%%%%%%%%%%%
\subsection{LWA Configuration Optimization}
\label{ssec:Problem}
Our goal is to leverage the frequency-dependent beamsteering performed by the \ac{lwa} to support high-rate data transfer in multi-user downlink communication systems. To that aim, we formulate the \ac{lwa}-based communication channel and characterize the relationship between the achievable rate and the transmitter's controllable parameters. %These parameters include the \ac{lwa} parameters $b,L$, the power allocation between spectral components of the input signal, and the choice of subbands. 
Specifically, 
\textcolor{NewColor}{we consider settings in which the configuration of the \ac{lwa}, i.e., the physical parameters $b$ and $L$, are tunable, either mechanically or electronically~\cite{lim2004metamaterial, chen2019electronically,javanbakht2021review}. The ability to tune these parameters, which necessitates a dedicated design,  
provides additional degrees of freedom for affecting the channel input-output relationship via the function $G(\cdot,\cdot)$ in \eqref{eq:channel input-output}.} 
\color{NewColor}
Under \ref{itm:OFDMA}, we focus on jointly tuning $b$ and $L$ along with  power $\{P_n\}_{n=1}^N$ as well as the subcarrier allocations $\{\mySet{S}_n \}_{n=1}^N$, based on the sum rate objective. 

To formulate the achievable sum rate performance for a given \ac{lwa} configuration, we define each $n$th subband channel vector, for all $K$ users as
$\myVec{h} _{n} \in \mathbb{C} ^{K}$. To encompass both \ref{itm:OFDM} and \ref{itm:OFDMA}, we have each entry $k\in\{1,\dots,K\}$ of $\myVec{h} _{n}$ be given by
\begin{align} 
\label{eq:fn-channel matrix}
\left[\myVec{h} _{n}\left(b,L,\mySet{S}_n\right)\right]_k \!=\! 
\begin{cases}
     G(\varphi _{k},f_{n} ; b,L) \PathLoss(\rho _{k}, f _{n}) & k\in \mySet{S}_n,\\
     0 &  k\notin \mySet{S}_n.
\end{cases}
\end{align}
Using this notation, the achievable sum rate performance at each frequency bin $f_n$ is given by the expression~\cite{shen2007sum}:
     \begin{align} \label{eq:1freq-rate}
    R_n\left(b,L,P _{n},\mySet{S}_n\right)
    \!=\!\frac{BW}{N}
    \log\left( 1\! + \! \frac{P_n\|\myVec{h}_{n}(b,L,\mySet{S}_n)\|^2}{BW\sigma ^{2}/N}\right).
    \end{align}
Accordingly, the average achievable sum rate of the wideband transmission is given by:
      \begin{align}\label{eq:sum rate}
    {R}\left(b,L,\{P _{n} \}_{n=1} ^{N},\{\mySet{S}_n\}_{n=1} ^{N} \right)\!=\! \sum_{n=1} ^N  R_n\left(b,L,P _{n},\mySet{S}_n\right).
    \end{align} 

The sum rate in \eqref{eq:sum rate} holds for both \ref{itm:OFDM} and \ref{itm:OFDMA}, depending on the setting of the spectral allocation in \eqref{eq: OFDMA_setting}. Clearly, the rate achieved when dividing the subbands between the users \ref{itm:OFDMA} is not larger than that achieved when having all users simultaneously utilize the entire band. 
\textcolor{NewColor}{However, in order to achieve the sum rate in \eqref{eq:sum rate} under \ref{itm:OFDM}, dedicated broadcast channel coding is required, see \cite[Ch. 8-9]{el2011network}. Still, the sum rate  represents an assessment of the quality of a wireless communication channels, that is widely adopted in studies dealing with the tuning parameters that affect the overall channel such as hybrid \ac{mimo}~\cite{thanh2023deep,lavi2023learn,qiao2020alternating}, metasuface antennas~\cite{shlezinger2021dynamic,wang2019dynamic}, and other antenna architectures~ \cite{abrardo2021intelligent}.} 

We aim to maximize the achievable sum rate performance under given physical and power constraints. The \ac{lwa} parameters $L$ and $b$ are typically constrained within some physical ranges, namely $b \in [b_{\min}, b_{\max}]$ and $L \in [L_{\min}, L_{\max}]$, to ensure a certain level of antenna efficiency. Based on the discussion above, the resulting problem of jointly tuning the \ac{lwa} and both power and frequency allocations can be mathematically expressed as follows:  
\begin{subequations}
\label{eq:rate optimization} 
  \begin{align} 
       &\hspace{-0.4cm} \mathop{\max}\limits_{b,L,\{P _{n}\}_{n=1} ^{N},\{\mySet{S}_n\}_{n=1} ^{N}} {R}(b,L,\{P _{n} \}_{n=1} ^{N},\{\mySet{S}_n\}_{n=1}^{N}) \\
   \text{s.t.}~ 
    &b_{\min}\leq b \leq b_{\max}, \label{eqn:b_constraint}\\
    &L_{\min}\leq  L \leq L_{\max}, \label{eqn:L_constraint}\\
    &\sum _{n=1}^N P_{n} \leq P, \label{eqn:power_sum}\\
    &P_n\geq0, \qquad \forall n\in\{1,\ldots,N\},\label{eqn:positive_power}\\
    &|\mySet{S}_n| \text{ follows \eqref{eq: OFDMA_setting}}, \quad  \forall n\in\{1,\dots,N\}\label{eqn:user_constraint}.
    \end{align}
    \end{subequations}
It can be seen that \eqref{eq:rate optimization} is a non-convex problem due to the complex dependence of the \ac{lwa}-induced channel on the optimization parameters, namely, the relationship induced by \eqref{eq:angular width} and \eqref{eq:fn-channel matrix} between the sum rate, the \ac{lwa} parameters $b$ and $L$, and the users allocations $\{\mySet{S}_n\}_{n=1}^N$. 
The first two constraints \eqref{eqn:b_constraint} and \eqref{eqn:L_constraint} over $b$ and $L$, respectively, ensure that the scales of the \ac{lwa} parameters satisfy the physical assumptions necessary for formulating the expressions in \eqref{eq:frequency-angle relation} and \eqref{eq:angular width}.
Then, \eqref{eqn:power_sum} and \eqref{eqn:positive_power} ensure that the power allocation withholds the total power constraint $P$, and that the optimized powers are non-negative values, as physically dictated. 
% Next, constraints \eqref{eqn:Const3}--\eqref{eqn:Const5} govern the division of the frequency band, ensuring that the optimized carrier frequencies are monotonically \textcolor{NewColor}{increasing},
% %non-decreasing,
% while supporting non-overlapping subbands that sum up to the entire available bandwidth. In addition, constraint \eqref{eqn:Const4} ensures that the allocated subbands withhold a physically compliant coherence width. 
Finally, \eqref{eqn:user_constraint} accommodates both \ref{itm:OFDM} and \ref{itm:OFDMA}, where under the latter,  \ac{ofdma} is achieved by allocating each subband solely to a single user.
\color{black}
%----------------------------------------------------------------------------------------
%	Downlink Beamforming
%----------------------------------------------------------------------------------------

\section{LWA-Aided Downlink Beamforming}
\label{sec:Beamforming}
In this section, we develop an alternating optimization framework for LWA-aided downlink beamforming in multi-user THz communication systems, \textcolor{NewColor}{and present the complexity of the resulting algorithm in terms of floating point operations}. Our framework is based on our characterization of the channel model and the resulting sum rate performance included in the previous Section~\ref{sec:System Model}.  We note that our design optimization problem \eqref{eq:rate optimization} is non-convex due to the complex dependence on  $b$, $L$, and $\{\mySet{S}_n\}_{n=1}^N$. However, when treating each set of optimization variables separately, the problems become simpler. Hence, we adopt an alternating optimization approach to set the tunable parameters based on the considered sum rate objective.
Specifically, we first focus on optimizing the \ac{lwa} configuration parameters, i.e., $b$ and $L$, as described in Subsection~\ref{ssec:LWAoptimization}. 
\textcolor{NewColor}{
Then, we discuss the setting of the power allocation $\{P _{n}\}_{n=1}^N$ in Subsection~\ref{ssec:Poweroptimization}, and how it is combined into subcarrier allocation $\{\mySet{S}_n \}_{n=1}^N$ for \ac{ofdma} in Subsection~\ref{ssec:subcarrier_optimization}.}
The overall algorithm is summarized in Subsection~\ref{ssec:optimization}, and its properties are discussed in Subsection~\ref{ssec:discussion}.

%In Subsection \ref{ssec:optimization}, we propose an alternating optimization algorithm to jointly optimize the LWA configuration and power allocation across subbands. Then, in Subsection \ref{ssec:discussion} we discuss the properties of this algorithm.

%%%%%%%%%%%%%%%%%%%%%%%%%%%
%%%	LWA Optimization
%%%%%%%%%%%%%%%%%%%%%%%%%%%
\subsection{LWA Configuration}
\label{ssec:LWAoptimization}
We start by optimizing the \ac{lwa} configuration $b$ and $L$ for fixed values of $\{P _{n} \}_{n=1}^N$ and $\{\mySet{S}_n\}_{n=1}^N$. 
We focus on the $i$th iteration of the overall alternating optimization, and thus, set the subband configuration parameters to be those obtained from the previous iteration, denoting them by $\{P _{n}^{(i-1)}, \mySet{S}_n^{(i-1)} \}_{n=1}^N$.

Since the sum rate objective is a non-convex function of $b$ and $L$, and both optimization decision variables are scalars taking value in some bounded range, we tune $b$ and $L$ for fixed $\{P _{n} \}_{n=1}^N$ and $\{\mySet{S}_n\}_{n=1}^N$ using a grid search. 
Specifically, we divide the ranges $[b_{\min},b_{\max}]$ and $[L_{\min},L_{\max}]$ into uniformly spaced grids, denoted by $\mySet{B}_{\rm grid}$ and $\mySet{L}_{\rm grid}$, respectively, such that the constraints in \eqref{eqn:b_constraint} and \eqref{eqn:L_constraint} are satisfied. 
By substituting these values in the optimization problem \eqref{eq:rate optimization}, all the constraints are redundant. Thus, we obtain that, at each $i$th iteration, one should scan all possible combinations on the grid to set as follows:
\begin{eqnarray}
\label{eq:b,L grid}
(b^{(i)},L^{(i)})=\hspace{6cm}\nonumber\\\mathop{\arg\max}\limits_{b,L \in \mySet{B}_{\rm grid}\times \mySet{L}_{\rm grid}}   {R}\left(b,L , \{P _{n}^{(i-1)} \}_{n=1}^N, \{\mySet{S}_n^{(i-1)}\}_{n=1}^N\right).
\end{eqnarray}
It can be seen that the complexity of solving \eqref{eq:b,L grid} depends on the grid resolution, as it involves $|\mySet{B}_{\rm grid}||\mySet{L}_{\rm grid}|$ ($|\cdot|$ returns the set cardinality) evaluations of the sum rate objective. 

% \textcolor{red}{theoretically, we can also compute the waterfilling here for each inspected grid point. Any reason why we opted no to do it?}

%%%%%%%%%%%%%%%%%%%%%%%%%%%
%%%	Power Optimization
%%%%%%%%%%%%%%%%%%%%%%%%%%%
\subsection{Spectral Power Allocation}
\label{ssec:Poweroptimization} 
We proceed to {the optimization of} the power allocation $\{P _{n} \}_{n=1}^N$. It can be seen from \eqref{eq:sum rate} that the sum rate objective is a convex function of  $\{P _{n} \}_{n=1}^N$, when $b$, $L$, and  $\{\mySet{S}_n\}_{n=1}^N$ are fixed. 
\color{NewColor}
Accordingly, we focus on the $i$th iteration of the overall alternating optimization algorithm, where the remaining optimization variables $b$, $L$, and  $\{\mySet{S}_n\}_{n=1}^N$ are set to $b^{(i)}$, $L^{(i)}$, and  $\{\mySet{S}_n^{(i-1)}\}_{n=1}^N$, respectively. By substituting these values into the optimization problem in \eqref{eq:rate optimization}, we obtain that, at each $i$th iteration, we need to set: 
\begin{align} 
\label{eq:PowerOpt}
\{P_n^{(i)}\}&=\mathop{\arg\max}_{\{P _{n}\}_{n=1} ^{N} }  {R}\left(b^{(i)},L^{(i)},  \{P _{n} \}, \{\mySet{S}_n^{(i-1)}\}\right)\nonumber\\
\text{s.t.} \quad
      &\sum _{n=1}^N P_{n} \leq P , \nonumber\\
      \quad&P_n\geq0, \qquad \forall n\in\{1,\ldots,N\}.
\end{align}
It can be easily concluded via \eqref{eq:sum rate} and the convexity of its constraints that \eqref{eq:PowerOpt} is a convex optimization problem, known to be solved via waterfilling~\cite[Ch. 5.3]{tse2005fundamentals}. The resulting optimal power allocation is thus the classical waterfilling solution, stated in the following lemma.
\begin{lemma}
\label{lem:Waterfilling}
    The optimal power allocation in \eqref{eq:PowerOpt} is given by
    \begin{equation} 
   \label{eq:WF Pn}
    P_n^{(i)}=\max\left\{\frac{1}{\nu} -  
    \frac{(BW/N)\cdot \sigma^2}{\|\myVec{h}_{n}(b^{(i)},L^{(i)},\mySet{S}_n^{(i-1)})\|^{2}}, 0\right\},
    \end{equation}
where $\nu>0$ is set such that $\sum_{n=1}^N P_n^{(i)} = P$. 
\end{lemma}
\color{black}
\smallskip
\begin{IEEEproof}
    The lemma is obtained from \cite[Sec. II-C]{he2013water}.
\end{IEEEproof}

The power allocation in \eqref{eq:WF Pn} is given in closed form. However, finding the exact setting of $\nu$ usually involves searching methods, with typical worst-case complexity order of $\mathcal{O}(N^2)$~\cite{he2013water,xing2020new}.

% For a fixed \ac{lwa} configuration, maximizing the sum rate with respect to $\{P _{n} \}_{n=1}^N$  and $\{\Delta f_n\}_{n=1}^N$ is still non-convex. However, for a given choice of $\{\Delta f_n\}_{n=1}^N$, the problem becomes convex and can be solved analytically with a solution similar to the conventional spectral power allocation of an evenly divided bandwidth, which is solved via waterfilling~\cite[Ch. 5.3]{tse2005fundamentals}.  Accordingly, once $b^{(i)}$ and $L^{(i)}$ are fixed, for a proposed choice of $\{\Delta f_n^{(i)}\}_{n=1}^N$ the optimal power allocation at the $i$th iteration is given by:

%%%%%%%%%%%%%%%%%%%%%%%%%%%
%%%	Subband Optimization
%%%%%%%%%%%%%%%%%%%%%%%%%%%
\subsection{Subcarrier Allocation}
\label{ssec:subcarrier_optimization}
\textcolor{NewColor}{When operating under \ref{itm:OFDM}, i.e., when each user utilizes all the \ac{ofdm} subbands, then the allocation variables $\{\mySet{S}_n\}$ are fixed via \eqref{eq: OFDMA_setting}. In such cases, setting is done by alternating between optimization via \eqref{eq:b,L grid} and via \eqref{eq:WF Pn}. However, under \ref{itm:OFDMA}, the subcarrier allocation can also be tuned}, 
\color{NewColor} 
while setting $(b,L)=(b^{(i)},L^{(i)})$ and $\{P_n=P_n^{(i)}\}_{n=1}^N$, as optimized in Subsections~\ref{ssec:LWAoptimization} and~\ref{ssec:Poweroptimization}, respectively.

In particular, under \ref{itm:OFDMA}, the spectral allocation variables $\{\mySet{S}_n\}_{n=1}^N$ can be represented as a sequence of $N$ indices, each in the finite set $\mySet{K}$. Thus, for a vector $\myVec{s}\in \mySet{K}^N$, we write the objective for fixed \ac{lwa} configuration and powers as
\begin{equation}\label{eq:discrete_optim_problem}
R^{(i)}(\myVec{s})\! \triangleq \!
   R\left(b^{(i)},L^{(i)},\{P_n^{(i)}\}_{n=1}^N, \{\mySet{S}_n=[\myVec{s}]_n\}_{n=1}^N\right), 
\end{equation}
where $R(\cdot)$ is given in~\eqref{eq:sum rate}. The fact that \eqref{eq:discrete_optim_problem} is formulated as objective function over a finite space, motivates its tackling using a \ac{ga} \cite{holland1992adaptation}. Such algorithms form an established discrete optimization framework that has also been  explored in the context of \ac{ofdma} \cite{chien2013optimizing,yang2010resource,ozbek2009adaptive}.
We thus hereby briefly review its main steps, which are summarized as Algorithm~\ref{alg:GA}. 

\begin{algorithm}
    \caption{Genetic Algorithm}\label{alg:GA}
    \textbf{Initialize:} Generate an initial population 
    \( \mathcal{P}_0 = \{\myVec{s}_p\}_{p=1}^{|\mySet{P}|} \)\;
    \For{$j = 0, \ldots, j_{\max}$}{
        For each \( \myVec{s} \in \mathcal{P}_j \), evaluate  \( R^{(i)}(\myVec{s}) \) in~\eqref{eq:discrete_optim_problem} (\ref{itm:fitness})
        \;    
        Set $\mySet{E}$ as top fitness scoring subset of $\mathcal{P}_j$ (\ref{itm:Elite})\;
        Select two parents based on the distribution in \eqref{eq:discrete_probability_distribution}\;
        From the parents, generate   crossover ($\mySet{C}$) and  mutation ($\mySet{M}$) children via \ref{itm:crossover} and \ref{itm:mutation}, respectively\;
        Update population $\mathcal{P}_{j}\to\mathcal{P}_{j+1}$ using \eqref{eq:population_update}\;        
    }    
    \Return Best solution found $\hat{\myVec{s}}$ in~\eqref{eq:GA_returned_value}\;
\end{algorithm}

Inspired by the principles of natural selection and genetics, \ac{ga} iteratively improves candidate solutions, with each iteration viewed as a generation. Specifically, the algorithm starts from an initial set of $|\mySet{P}|$ candidate allocations (termed population), denoted \( \mathcal{P}_0 = \{\myVec{s}_p\}_{p=1}^{|\mySet{P}|} \). 
Each candidate solution  $\myVec{s}_p$ is randomized uniformly   over $\mySet{K}^N$. At each iteration on index $j$, the \ac{ga} updates the current population $\mySet{P}_j$ into $\mySet{P}_{j+1}$ by forming three sets: {\em elite individuals} $\mySet{E}$; {\em crossover children} $\mySet{C}$; and {\em mutation children} $\mySet{M}$. These sets are obtained via the following steps~\cite{lim2017crossover}: 

\begin{enumerate}[label=\textbf{G\arabic*}, wide=0pt, leftmargin=*]  
\item \label{itm:fitness}{\bf Fitness Evaluation.} 
Each solution \( \myVec{s}_p  \in \mySet{P}_j\) is evaluated on the objective function $R^{(i)}(\myVec{s}_p)$ in~\eqref{eq:discrete_optim_problem} via~\eqref{eq:sum rate}.

\item \label{itm:Elite} {\bf Elite individuals ($\mySet{E}$).} 
The ${|\mySet{E}|}$ solutions with the best fitness values in the current generation, automatically survive to the next generation.

\item \label{itm:Parents} {\bf Parents Selection.}
Then, each scoring $R^{(i)}(\myVec{s}_p)$ is being normalized to form a discrete probability distribution:
\begin{equation}\label{eq:discrete_probability_distribution}
     P(\myVec{s}_p) \triangleq \frac{R^{(i)}(\myVec{s}_p)}{\sum_{j=1}^{P} R^{(i)}(\myVec{s}_p)}, \quad p = 1, 2, \dots, |\mySet{P}|.
\end{equation}
According to \eqref{eq:discrete_probability_distribution}, two `parents` $\myVec{s}_{p_1}$ and $\myVec{s}_{p_2}$ are sampled (with replacement) from the population $\mySet{P}_j$.

\item \label{itm:crossover} {\bf Crossover Children ($\mySet{C}$).}
For a fixed crossover fraction $\rho \in [0,1]$, 
a total of $|\mySet{C}| = \rho \left(|\mathcal{P}|-|\mySet{E}|\right)$ new candidate allocations ('children') are created by combining the vectors of the pair of parents. That is, the entries of each child vector $\myVec{s}^{\rm Cross}_c \in \mySet{C}$ are randomly selected to be the same coordinate from one of the two parents:
\begin{align}\label{eq:crossover_children}
    [\myVec{s}^{\rm Cross}_c]_n \triangleq\begin{cases}
        [\myVec{s}_{p_1}]_n & \text{w.p } 0.5\\
        [\myVec{s}_{p_2}]_n & \text{w.p }  0.5
    \end{cases},\quad n = 1, \ldots, N.
\end{align}

\item \label{itm:mutation} {\bf Mutation Children ($\mySet{M}$).}
The remaining candidate solutions, termed `mutation children`, are created by randomly changing the parents genes. Specifically, each $\myVec{s}^{\rm Mutation}_c \in \mySet{M}$ is obtained by adding Gaussian noise to a parent $\myVec{s}_{p_i}, \ i\in\{1,2\}$; with a subsequent rounding for assuring  elements in $\mySet{K}^N$.  

\item {\bf Population Update.} The generated generation is obtained as follows:
\begin{align}\label{eq:population_update}
    \mySet{P}_{j+1} =  \mySet{E} \cup \mySet{C} \cup \mySet{M}.
\end{align}
\end{enumerate}

The process repeats until a maximum number of $j_{\max}$ generations, and returns
\begin{equation}\label{eq:GA_returned_value}
    \hat{\myVec{s}} = {\arg\max}_{\myVec{s}\in\mySet{P}_{j_{\max}}} R^{(i)}(\myVec{s}).
\end{equation}
It is noted that both processes of Mutation and Crossover are essential to a \ac{ga}. Crossover enables the extraction of the best genes from different individuals and recombine them into potentially superior children. Mutation adds to the diversity of a population and thereby increases the likelihood that the algorithm will generate individuals with better fitness values \cite{lim2017crossover}.

The complexity of the \ac{ga} in Algorithm~\ref{alg:GA} depends on several factors, including the number of iterations (generations), denoted by $j_{\max}$, the population size $|\mySet{P}|$; as well as, the crossover and mutation functions, and, finally, the fitness evaluation—which is in the order of $\mathcal{O}(N)$ according to \eqref{eq:sum rate}. As a result, the wort-case complexity is $\mathcal{O}(j_{\max} |\mySet{P}| N)$.

\color{black}
%%%%%%%%%%%%%%%%%%%%%%%%%%%
%%%	Alternating Optimization
%%%%%%%%%%%%%%%%%%%%%%%%%%%
\subsection{\textcolor{NewColor}{Overall
Alternating Optimization for LWA Beamforming}}
\label{ssec:optimization}
\textcolor{NewColor}{In this subsection, we present the overall algorithm for solving~\eqref{eq:rate optimization}, using the proposed alternating optimization framework. }
The individual tools for setting each of the optimization variables dictating the beampattern achieved with the proposed \ac{lwa}-aided downlink communications are combined into an overall beamforming algorithm based on alternating optimization. 
\color{NewColor}
We identify differences in the basic characteristics of the problems' parameters that can be divided into two groups: 
\textit{i}) $b$ and $L$ are the physical scales of the \ac{lwa}, hence, both share similar constraints in \eqref{eq:rate optimization}; \textit{ii}) The constraints on the power and subcarrier allocation, $\{P_n\}_{n=1}^N$ and $\{\mySet{S}_n\}_{n=1}^N$, respectively, are
derived from the applied communication model. 
Consequently, our proposed algorithm alternates between three main steps: 
\textit{i}) Optimizing the \ac{lwa} configuration $b$ and $L$ for fixed $\{P _{n} \}_{n=1}^N$ and $\{\mySet{S}_n\}_{n=1}^N$ (as detailed in Subsection~\ref{ssec:LWAoptimization}); 
\textit{ii}) setting the power allocation $\{P _{n} \}_{n=1}^N$ for fixed \ac{lwa} $b$ and $L$ and subcarrier allocation (based on the closed-form waterfilling approach in Subsection~\ref{ssec:Poweroptimization});
and \textit{iii}) when operating under \ref{itm:OFDMA}, setting the subcarrier allocation $\{\mySet{S}_n\}_{n=1}^N$ for fixed \ac{lwa} $b$ and $L$ and power allocation (based on the \ac{ga} algorithm in Subsection~\ref{ssec:subcarrier_optimization}).
The overall procedure for solving~\eqref{eq:rate optimization} is summarized in Algorithm~\ref{alg:LWA opt}.
 
\begin{algorithm}
    \caption{Proposed \ac{lwa} Beamforming}
    \label{alg:LWA opt}     \SetKwInOut{Input}{Input}  
    \Input{Users positions $\{\varphi_k,\rho_k\}_{k=1}^K$; total power $P$;  
    total frequencies $N$; bandwidth $BW$.} 
    \SetKwInOut{Initialization}{Initialize}
    \Initialization{\textcolor{NewColor}{Power allocation $P _{n}^{(0)}= \frac{P}{N}$};\\
   % \ac{ga}'s number of generations $G$, population size $P$, and mutation probability $p_m$;
    iterations limit $i_{\max}$.}
  %  \Output{The learned parameters $\bm{\mu}^{(o)}$}
        \For{$i = 1, \ldots, i_{\max}$}{
        \label{stp:Step1}
        \textbf{\textcolor{NewColor}{Step 1: Set LWA Configuration Parameters}} \hspace{2cm}\\
                     Grid search for $(b^{(i)},L^{(i)})$ via \eqref{eq:b,L grid}\;     
         \label{stp:Step2}            
         \textbf{\textcolor{NewColor}{Step 2: Set Power Allocation}} \hspace{-0.15cm}\\
                    Set $\{P_n^{(i)}\}_{n=1}^N$ via~\eqref{eq:WF Pn}\;            
        \label{stp:Step3}            
         \textbf{\textcolor{NewColor}{Step 3: Set Subcarrier Allocation (under \ref{itm:OFDMA})}} \hspace{-0.15cm}\\
                    Set $\{\mySet{S}_n^{(i)}\}_{n=1}^N$ via Algorithm~\ref{alg:GA}\;            
        }
        \KwRet{$b^{(i_{\max})}, L^{(i_{\max})}, \{P_n^{(i_{\max})}\}_{n=1}^N, \{\mySet{S}_n^{(i_{\max})}\}_{n=1}^N$.}
\end{algorithm}

{\bf Complexity:} We note that Algorithm~\ref{alg:LWA opt} iterates $i_{\max}$ times between grid searching over \eqref{eq:b,L grid}, waterfilling over \eqref{eq:WF Pn}, and applying Algorithm~\ref{alg:GA}. As detailed in Subsection~\ref{ssec:LWAoptimization}, each respectively involves (worst-case) complexity order of $\mathcal{O}(|\mySet{B}_{\rm grid}| |\mySet{L}_{\rm grid}|),\ \mathcal{O}(N^2)$, and $\mathcal{O}(j_{\max}|\mySet{P}| N)$. 
Accordingly, assuming that the \ac{ga} hyperparameters $j_{\max} |\mySet{P}|$ are of an order not surpassing $\mySet{O}(N)$, the overall complexity  of Algorithm~\ref{alg:LWA opt} is of the order 
\begin{equation}
\label{eqn:Complexity}
     \mathcal{O}\left(i_{\max} \left( |\mySet{B}_{\rm grid}||\mySet{L}_{\rm grid}| +N^2 \right) \right).
\end{equation}
% The complexity order in \eqref{eqn:Complexity} consists of two summands, representing the two main stages of Algorithm~\ref{alg:LWA opt}. 
The identity of the dominant term depends on the grid sizes. 

\textcolor{NewColor}{The overall computational overhead can thus be balanced by tuning the optimization parameters, and particularly the grid sizes and iterations, potentially trading performance for complexity. Alternatively, one can consider integrating machine learning based computations~\cite{zappone2019wireless}, possibly adopting paradigms for combining iterative optimization with deep learning, e.g., deep unfolding~\cite{shlezinger2022model}, leveraging data to enhance rapid operation. }

\color{black}
%%%%%%%%%%%%%%%%%%%%%%%%%%%
%%%	Discussion
%%%%%%%%%%%%%%%%%%%%%%%%%%%
\subsection{Discussion}
\label{ssec:discussion}
Our proposed physically compliant modeling of \ac{lwa}-based wireless communications facilitates unveiling the potential of this antenna technology for future high-frequency wireless systems. From a hardware perspective, the \ac{lwa} technology has notable gains in terms of cost and power consumption due to its ability to steer wideband THz beams in a controllable manner with a single element. This approach eliminates the need for numerous wideband radio frequency chains and analog beamformers, which become extremely costly at THz frequencies and are typically employed by conventional massive \ac{mimo} architectures. 
\textcolor{NewColor}{While we focus on \acp{lwa} with a single slit, which implements 2D beamforming, one can also consider \acp{lwa} with multiple slits, that can steer beams in both azimuth and elevation~\cite{yeh2023security}. Thus, our framework can be extended to settings where users are at different elevation angles, with proper adaptation of the channel model. The angular-frequency coupling of \ac{lwa}, which may be viewed as limiting the ability to obtain ultra-wideband communications with a single user (though it can still provide GHz-scale bandwidth for a given angular direction), is exploited in our framework for supporting multiple users.}

The proposed alternating optimization method in Algorithm~\ref{alg:LWA opt} for solving~\eqref{eq:rate optimization} provides a means to tune \acp{lwa} to achieve desirable beampatterns that support multiple users. This capability is numerically demonstrated in the following Section~\ref{sec:Experimental Study}. We note that the computational burden of tuning the \ac{lwa} via Algorithm~\ref{alg:LWA opt} is dictated by the grid sizes, the number of iterations, and the complexity of the extended waterfilling approach. The former two factors are design hyperparameters, which can be configured based on a desired complexity level. Computing \eqref{eq:WF Pn} involves an iterative search, whose complexity grows quadratically with $N$  in the worst case, and often grows only linearly with $N$ (depending on how many subchannels are active)~\cite{he2013water}.   

\textcolor{NewColor}{Our optimization objective in \eqref{eq:rate optimization} is an evaluation of the quality of the wireless channel. Under \ref{itm:OFDM}, it does not assume any orthogonality imposed, allowing multiples messages to be superimposed on the same beam. As such, it does not necessarily degrade in the presence of closely separated users. Still, we can identify the effect of employing \acp{lwa} when distinct user-wise beams are desired from their induced channel model characterized in Section~\ref{sec:System Model}. There, one  readily observes that the ability to generate narrow beams per user varies between different angles, where users whose angular position grows experience wider beams (with lesser angular separation) compared to users with lesser relative angles.}

In our proposed Algorithm~\ref{alg:LWA opt}, we opted 
\textcolor{NewColor}{to optimize} the power allocation using waterfilling \textcolor{NewColor}{in a separate step (Step~\ref{stp:Step1}), rather than integrating it directly into} the tuning of the LWA physical parameters $b$ and $L$ \textcolor{NewColor}{(Step~\ref{stp:Step2}).  This decision is based on two main considerations. First, while} %Similar to the optimization of the spectral division, 
tuning the \ac{lwa} parameters impacts the channel representation,
%based on which the power allocation was optimized. However, we perceive 
the coupling between the \ac{lwa} parameters and the optimal power allocation is relatively weak, \textcolor{NewColor}{in the sense that the variations in the power allocations due to different settings of $(b,L)$ is often experimentally shown to be minor. Second,} the grid sizes for the search of $b$ and $L$ are notably larger than the grid size of the spectral width of each subband, and of the number of frequencies $N$. \textcolor{NewColor}{Including the waterfilling in the tuning of $b$ and $L$ would thus introduce} additional computational complexity \textcolor{NewColor}{with limited expected impact on their optimal configuration.  Therefore, we focus on alternating between the considered variables for efficient optimization, while leaving}
 the impact of tuning the power allocation simultaneously with $b$ and $L$ for future investigation.  

% \textcolor{red}{TODO YAELA - add two sentences here justifying why we opt not to include WF in tuning of $b,L$ -- clarify that it is expected to less impact the tuning of these parameters compared to the frequency widths, and that it comes at the cost of increased complexity. Leave for future investigation.}

We finally note that the \ac{lwa} communication model in \eqref{eq:channel input-output} is formulated for a generic path loss function $\PathLoss(\cdot,\cdot)$. Thus, while we set it to be range-dependent for simplicity, one can also use frequency-dependent profiles, which are expected to be encountered in THz. 
\textcolor{NewColor}{Another potential extension of our work is the analysis of the uplink, where a multiple-access channel model based on \eqref{eq:sum rate} is expected to be representative due to reciprocity, while the sum rate formulation differs from the one considered here~\cite[Ch. 6]{tse2005fundamentals}.}
Moreover, while our beampattern design is based on the sum rate maximization in \eqref{eq:rate optimization}, alternative relevant problems that follow from the \ac{lwa} model can be considered. For instance,  
we currently focus on the upper bound for the achievable sum rate calculated by \eqref{eq:sum rate}, \textcolor{NewColor}{and thus do not account for fairness or user-specific data rate requirements, which require a modification of the objective}. 
Further, \textcolor{NewColor}{inspired by the importance of having controllable beampatterns in \ac{mimo}~\cite[Ch. 10]{tse2005fundamentals} and hybrid \ac{mimo}~\cite{shlezinger2023ai} systems, we consider a framework where one can adjust the parameters $b$ and $L$ that affect the beampattern.}
In practice, these \ac{lwa} configuration parameters are controlled via \textcolor{NewColor}{electronic or} mechanical measures, which may constrain how frequently they can be modified. Finally, our formulation assumes that the users' locations are known, without exploring how they are recovered by an \ac{lwa}-aided \ac{bs}. However,  recently, \cite{boljanovic2023joint} have shown that employing THz rainbow transmissions can in fact facilitate acquiring such localization estimates, which can be potentially combined with our subsequent beamforming task. These extensions of our study, whose aim is to introduce the framework of \ac{lwa}-aided communications and its potential for THz systems, are left for future work.

%----------------------------------------------------------------------------------------
%	Experimental Study
%----------------------------------------------------------------------------------------
\section{Numerical Results}
\label{sec:Experimental Study}
This section numerically evaluates \ac{lwa}-equipped \acp{bs} in supporting multiple downlink users via our proposed beamforming design\footnote{The source code and hyperparameters used in this experimental study is available at: \url{https://github.com/langnatalie/LWA}.}. 
\textcolor{NewColor}{\textcolor{NewColor}{We begin by formulating the setting used in our simulations in Subsection~\ref{ssec:setting}}, after which we divide our numerical study into two parts: 
The first, reported in Subsection~\ref{ssec:beampatterns}, evaluates the gains of 
\textcolor{NewColor}{\ac{lwa}-aided \ac{ofdm} and \ac{ofdma} optimization, illustrating the beam beampatterns and assessing the sum rate achieved using a single \ac{lwa}.}
The second part, comprised of Subsection~\ref{ssec:rates_compare}, focuses on evaluating the sum rate achieved with \acp{lwa}, and comparing it to alternative wideband THz antennas, including (extremely costly) fully digital \ac{mimo} signaling, and hybrid analog/digital architectures.} 

\subsection{Experimental Setting}
\label{ssec:setting}
To evaluate the capability of an \ac{lwa}-aided \ac{bs} in generating directional beams, we consider the channel model detailed in Section \ref{sec:System Model} in the frequency range of $[0.2,0.8]$ THz. This range is chosen due to
\textcolor{NewColor}{its relevance to the operating characteristics of \acp{lwa} at THz frequencies, where wideband operation is feasible, and aligns with similar wideband \acp{lwa} experiments—albeit at lower frequencies—reported in~\cite{ghasempour2020single, ghasempour2020leakytrack} and \cite{kludze2024frequency}. As we focus on relatively short range distances (below 1km), atmospheric attenuation is neglected~\cite{Yang_2011}.}

The band is divided into $N=150$ frequency bins. We randomize $K=4$ users for small and $K=16$ users for large scale scenarios; with relative angles and distances drawn uniformly in the ranges $[10,55]^{\circ}$ and $[10,20]$ meters, respectively. For each setting, we use the proposed Algorithm~\ref{alg:LWA opt} to tune the \ac{lwa} configuration within the ranges $[b_{\min},b_{\max}]= [0.9,1.1]$ mm and $[L_{\min},L_{\max}]= [10,\textcolor{NewColor}{30}]$ mm, where  the total power constraint is $10\times$ the average noise power, i.e., $P=10\sigma^2\frac{BW}{N}$.  
\textcolor{NewColor}{We deployed the \ac{ga} Algorithm~\ref{alg:GA} with population size $|\mySet{P}|=20$, elite individuals $|\mySet{E}|=2$, crossover fraction  $\rho=0.8$, and a maximal number of generations $j_{\rm max}=20$ .}

\textcolor{NewColor}{To calculate the sum rate (in bits-per-second), $30$ Monte-Carlo simulations were averaged, each randomly chooses the locations of the users. We define the \ac{snr} as ${\rm SNR} = \frac{P}{N\sigma^2}$ per subcarrier (as the noise is white). Unless stated otherwise, we set the \ac{snr} to $0$ dB.}

\subsection{LWA-aided OFDM and OFDMA Optimization}
\label{ssec:beampatterns}

\subsubsection{Beampattern Comparison}
\textcolor{NewColor}{We begin with a small-scale multi-user scenario, setting $K=4$. We illustrate in Figs.~\ref{fig:beams_OFDM}-\ref{fig:beams_OFDMA} two different beampatterns}, i.e., the energy radiated towards each position $(\varphi, \rho)$ over the entire spectrum \textcolor{NewColor}{under \ac{ofdm} \ref{itm:OFDM} and \ac{ofdma} \ref{itm:OFDMA} signaling, respectively.} The energy at each position is computed in logscale as $\log(\sum_{n=1}^N |G(\varphi,f_n) \PathLoss (\rho, f_n)|^2 P_n)$ using our proposed Algorithm~\ref{alg:LWA opt}. For simplicity, we  set the attenuation coefficient to scale as $\PathLoss(\rho,f ) \propto 1/{\rho}$. 

\begin{figure*}
		\centering
        \begin{subfigure}{0.48\textwidth}
            \includegraphics[width=0.9\linewidth]{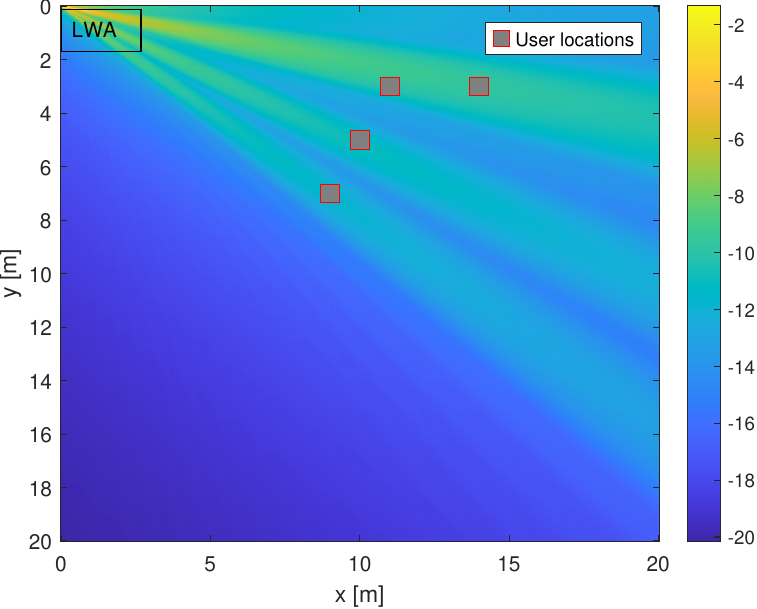}
         \caption{Directed beams towards all $K=4$ users.} % good_beams OFDM
         \label{fig:good_ofdm}
         \end{subfigure}
         \hfill         
        \begin{subfigure}{0.48\textwidth}
            \includegraphics[width=0.9\linewidth]{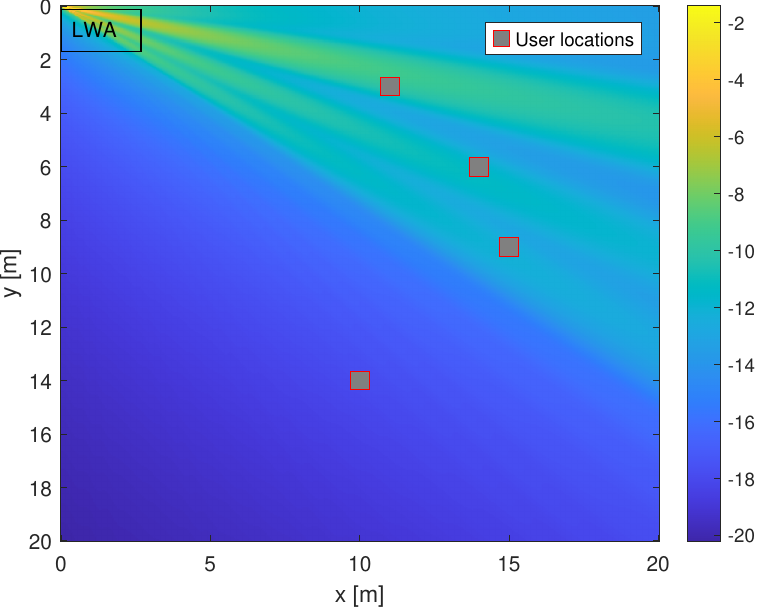}
         \caption{Directed beams towards $3$ out of $K=4$ users.} % bad_beams OFDM
         \label{fig:bad_ofdm}
        \end{subfigure}
        \vspace{5mm}
    \caption{Beampatterns achieved using a single \ac{lwa} with \ac{ofdm} \ref{itm:OFDM}, $K=4$. The coordinate system is compliant with Fig.~\ref{fig:Channel}.} 
    \label{fig:beams_OFDM}
\end{figure*} 

\begin{figure*}
		\centering
        \begin{subfigure}{0.48\textwidth}
         \includegraphics[width=0.9\linewidth]{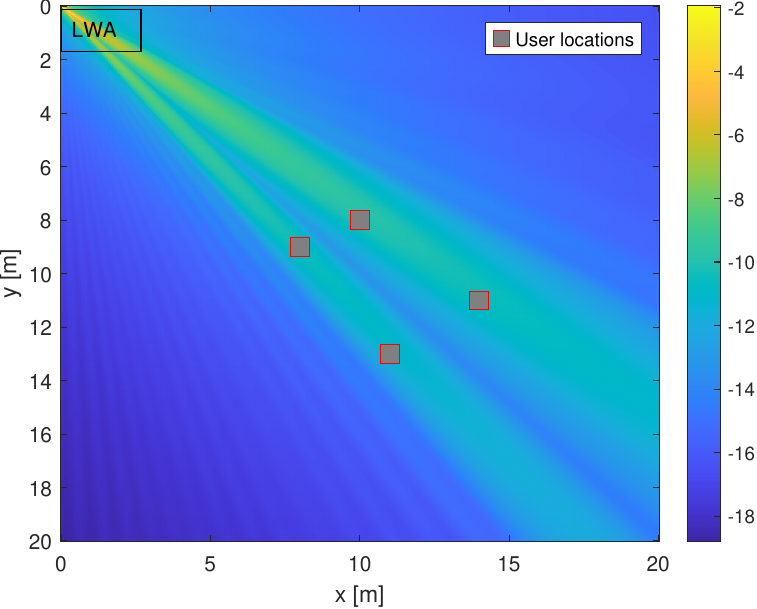} % good_beams OFDMA
         \caption{Directed beams towards all $K=4$ users.}
         \label{fig:good_ofdma}  
         \end{subfigure}
         \hfill
         \begin{subfigure}{0.48\textwidth}
         \includegraphics[width=0.9\linewidth]{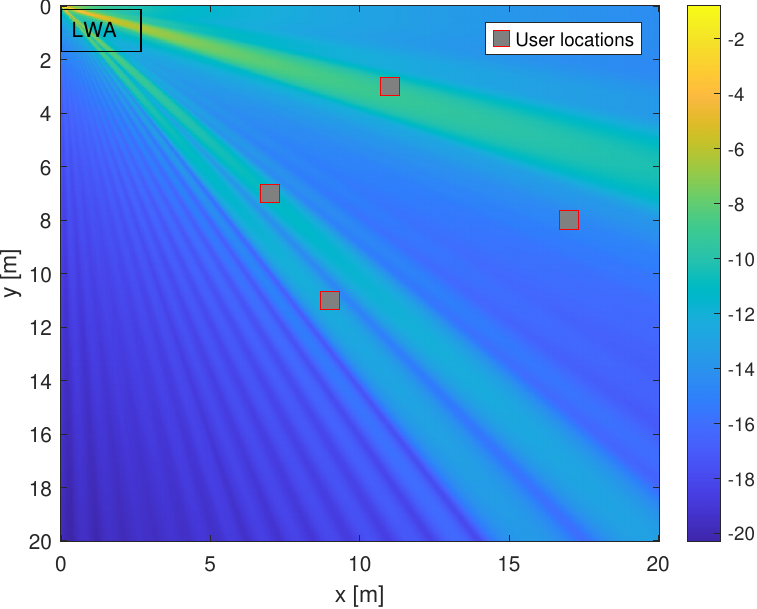} % bad_beams OFDMA
         \caption{Directed beams towards $3$ out of $K=4$ users.}
         \label{fig:bad_ofdma}  
     \end{subfigure}
     \vspace{5mm}
    \caption{Beampatterns achieved using a single \ac{lwa} with \ac{ofdma} \ref{itm:OFDMA}, $K=4$. The coordinate system is compliant with Fig.~\ref{fig:Channel}.} 
    \label{fig:beams_OFDMA}
\end{figure*} 

It is observed in Figs.~\ref{fig:beams_OFDM}-\ref{fig:beams_OFDMA} that a single \ac{lwa} can generate relatively directed beams. Moreover,  Figs.~\ref{fig:beams_OFDM}-\ref{fig:beams_OFDMA} indicate on the inherent lack of symmetry of the \ac{lwa}-based channel towards user locations/frequencies. Specifically, as the  \ac{aod} changes monotonically with the frequency by \eqref{eq:frequency-angle relation}, users located at small angles with respect to the \ac{lwa} axis communicate at higher frequencies (with transmission power spread over multiple bins), unlike those located at larger angles. By inspecting \eqref{eq:angular width}, it is also evident that a monochromatic beam's width is dependent on the beam's frequency, as $\beta$ and $k_0$ are functions of $f$, and thus, higher frequency components are steered into narrower and more accurate beams. Consequently, for a system with users of similar distance from the \ac{lwa} and equal power allocation for all engaging frequency components, the \ac{lwa}-induced channel can naturally achieve higher rates communicating with users at smaller angles (and higher frequencies). 
Accordingly, the proposed Algorithm~\ref{alg:LWA opt} is prone to assign more resources to steering beams towards the users with smaller angles. This can result in a relatively balanced service of multiple users, as in Figs.~\ref{fig:good_ofdm}-\ref{fig:good_ofdma}, or alternatively, steering most of the signal power towards users at lower angles, as in Figs.~\ref{fig:bad_ofdm}-\ref{fig:bad_ofdma}. These beampatterns show the ability of a \ac{lwa} to generate directed beams, and also indicate that in settings where it is crucial to guarantee balanced service of multiple users, alternative objectives, e.g., minimal rate, should be considered. 

We proceed to showcasing that the above insights generalize to to large-scale scenarios, considering a setup with  $K=16$ users. Similar to the small-scale scenario, we first visualize candidate beampatterns achieved in Figs.~\ref{fig:beams_OFDM_large-scale}-\ref{fig:beams_OFDMA_large-scale}. In these figures we note that the \ac{lwa}-based channel, set according to our proposed Algorithm \ref{alg:LWA opt}, can support relatively directed beams towards given user locations. Once again, we witness that some settings are matched with a solution that steers beams towards all $K=16$ user locations, as demonstrated in Figs.~\ref{fig:good_ofdm_large-scale},-\ref{fig:good_ofdma_large-scale} while in other cases the beams are steered only towards a partial group of users, with notable preference towards users located at small angles, as demonstrated in  Fig. \ref{fig:bad_ofdm_large-scale}-\ref{fig:bad_ofdma_large-scale}.  These results also demonstrate the benefits of leveraging the dispersed angular range, facilitated by the THz rainbow that is created by \acp{lwa}, in supporting a large number of users using a single transmit antenna element. 

\begin{figure*}
		\centering
        \begin{subfigure}{0.48\textwidth}
            \includegraphics[width=0.9\linewidth]{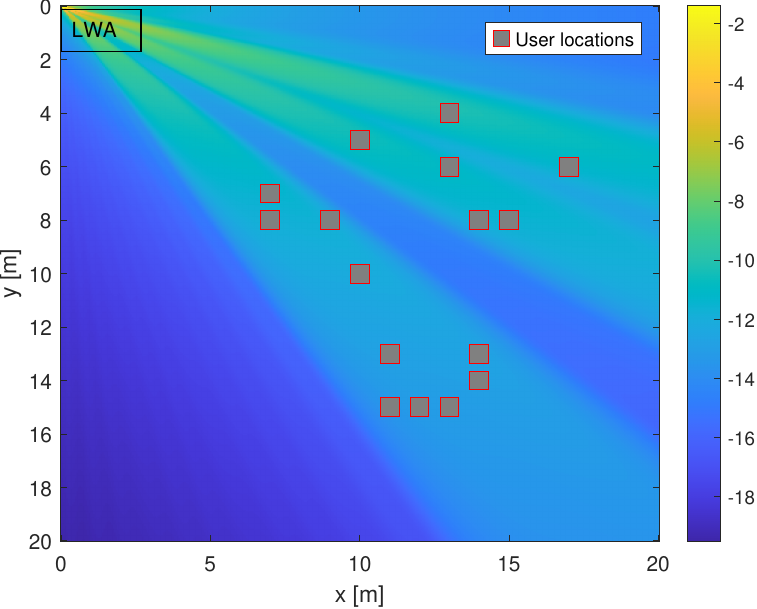}
         \caption{Directed beams towards all $K=16$ users.} % good_beams OFDM
         \label{fig:good_ofdm_large-scale}
         \end{subfigure}
         \hfill         
        \begin{subfigure}{0.48\textwidth}
            \includegraphics[width=0.9\linewidth]{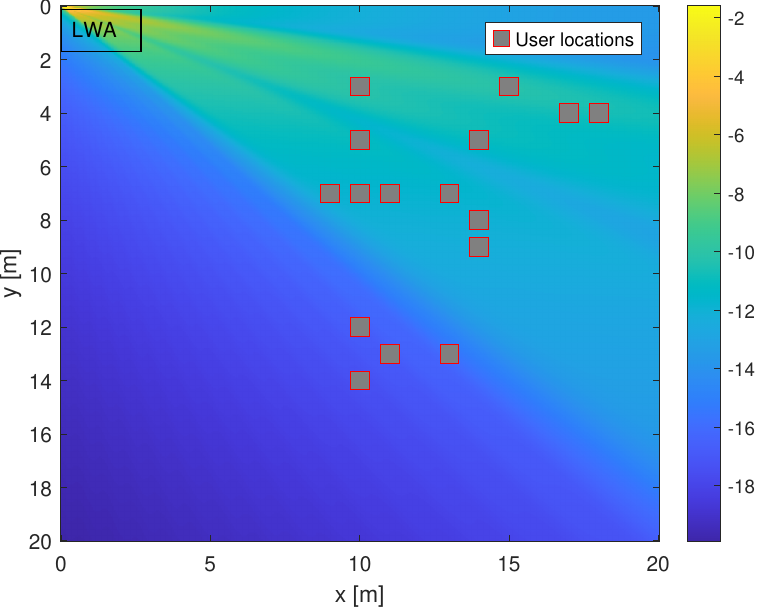}
         \caption{Directed beams towards $12$ out of $K=16$ users.} % bad_beams OFDM
         \label{fig:bad_ofdm_large-scale}
        \end{subfigure}
        \vspace{5mm}
    \caption{Beampatterns achieved using a single \ac{lwa} with \ac{ofdm} \ref{itm:OFDM}, $K=16$.  The coordinate system is compliant with Fig.~\ref{fig:Channel}.} 
    \label{fig:beams_OFDM_large-scale}
\end{figure*} 

\begin{figure*}
		\centering
        \begin{subfigure}{0.48\textwidth}
         \includegraphics[width=0.9\linewidth]{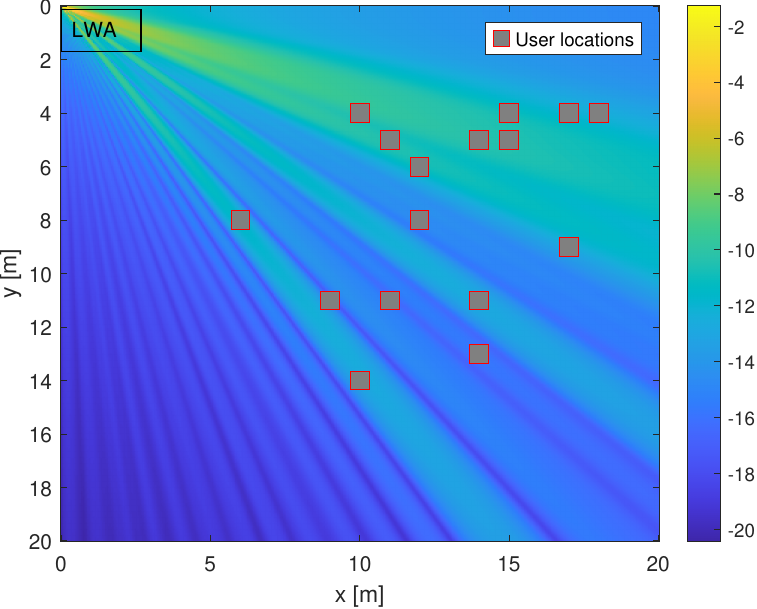} % good_beams OFDMA
         \caption{Directed beams towards all $K=16$ users.}
         \label{fig:good_ofdma_large-scale}  
         \end{subfigure}
         \hfill
         \begin{subfigure}{0.48\textwidth}
         \includegraphics[width=0.9\linewidth]{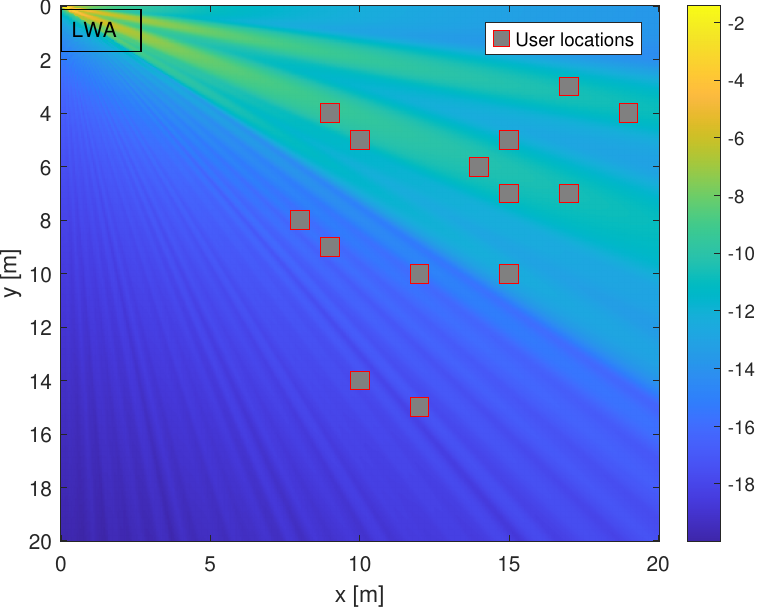} % bad_beams OFDMA
         \caption{Directed beams towards $11$ out of $K=16$ users.}
         \label{fig:bad_ofdma_large-scale}  
     \end{subfigure}
     \vspace{5mm}
    \caption{Beampatterns achieved using a single \ac{lwa} with \ac{ofdma}  \ref{itm:OFDMA}, $K=16$. The coordinate system is compliant with Fig.~\ref{fig:Channel}.} 
    \label{fig:beams_OFDMA_large-scale}
\end{figure*} 

\subsubsection{Power and Subband Allocation}
\label{subsec:SimBand}
As our optimization framework supports tuning power division and subband allocation, 
%The formulation of the achievable sum rate performance in a manner which is not confined to \ac{ofdm}/A builds on the assumption that each subband width does not exceed its corresponding coherence bandwidth $d$. For this purpose, one has to limit the allocated subbands' widths to ensure that each frequency subband can be well represented by computing~\eqref{eq:angular width} and~\eqref{eq:fn-channel matrix}, and that the channel behavior is approximately constant for all frequencies under the same subband. To ensure that this assumption does hold, 
we proceed visualize the sub-channel norms calculated for the \ac{lwa}-optimized channel using $K=4$ users. 
We begin by depicting the obtained power allocation and corresponding (normalized) channel norms when using \ac{ofdm} signaling in Fig.~\ref{fig:analysis_OFDM}. We note that the resulting plot for the sub-channel norms portray piece-wise-constant curves that can be considered as good approximations to the channel representation achieved under the \ac{ofdm} setting. 
%We, thus, conclude that the choice $d=BW/40$ used in our simulation study is a tight enough constraint over the coherence width. Tighter constraints would naturally, keep the physical demands of the system. However, we advise not to choose an overly loose constraint under the given setting since the results may no longer represent physically achievable sum rate performance.

\begin{figure*}
		\centering
        \begin{subfigure}{0.48\textwidth} % OFDM
         \includegraphics[width=0.9\linewidth]{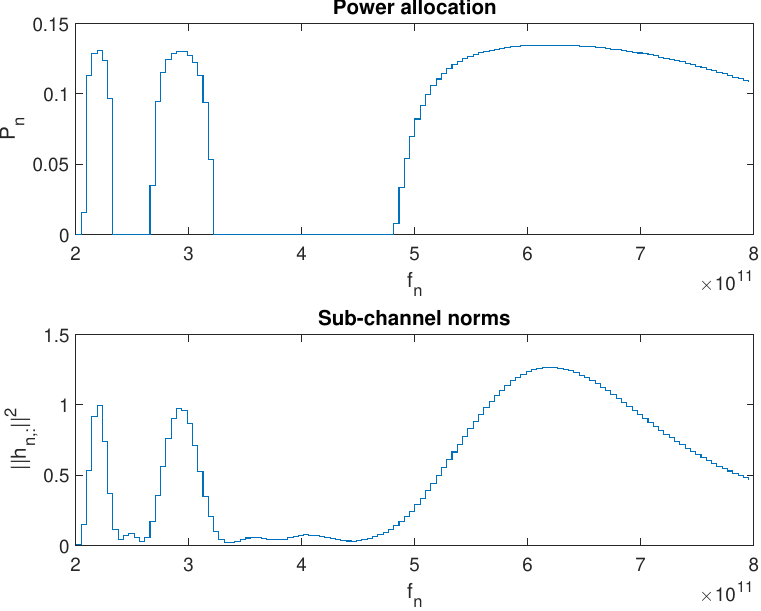} 
         \caption{Setting \ref{itm:OFDM}.}
        \label{fig:analysis_OFDM}  
         \end{subfigure}
         \hfill
         \begin{subfigure}{0.51\textwidth}  % OFDMA
         \includegraphics[width=0.9\linewidth]{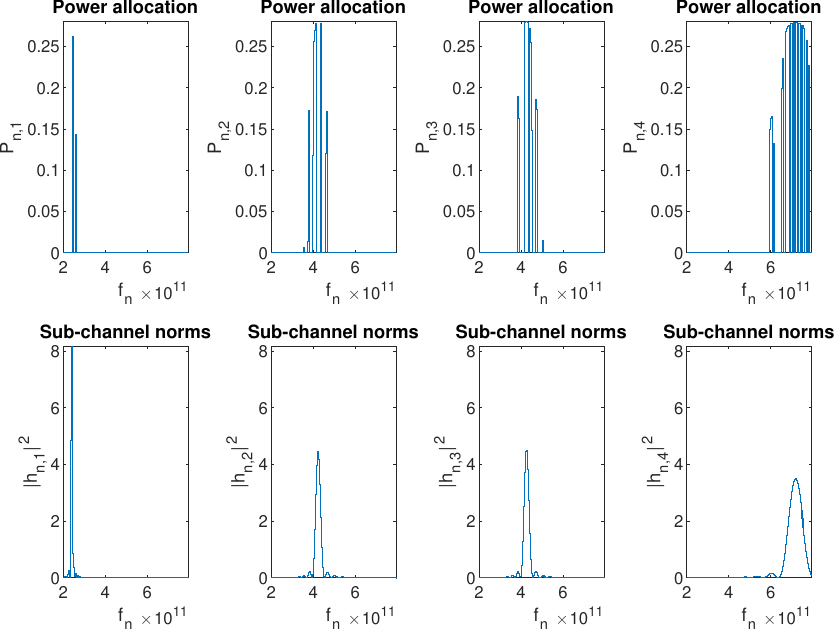}
        \caption{Setting \ref{itm:OFDMA}.}
        \label{fig:analysis_OFDMA}  
     \end{subfigure}
     \vspace{5mm}
    \caption{Power allocations and corresponding channel norm with $K=4$ users.} 
\end{figure*} 

\textcolor{NewColor}{In addition to the channel norm plots, we depict in Fig.~\ref{fig:analysis_OFDMA} the users allocation achieved under \ref{itm:OFDMA} with Algorithm~\ref{alg:GA}, as well as the corresponding power allocations. 
Our results demonstrate that, when allowing the usage of simple coding schemes utilized in standard orthogonal systems, via, e.g., \ac{ofdma}, the \ac{lwa} signaling effectively exploits, for each user, the sub-channels corresponding to valuable frequencies with large channel norms.} 
Hence, the dedicated alternating optimization method in Algorithm~\ref{alg:LWA opt} enables to properly exploit the channel characteristics induced by the considered \ac{lwa} technology. % without violating the underlying assumption of approximately flat sub-channels.

\subsection{Achievable sum rate Performance}
\label{ssec:rates_compare}
We proceed to evaluating the sum rate achieved with \acp{lwa}. We first assess the rate gap due to the usage of practically simpler subband division \ref{itm:OFDMA}, after which we compare performance with costly \ac{mimo}-based \acp{bs}. 

\subsubsection{LWA-based OFDM \ref{itm:OFDM} vs. OFDMA \ref{itm:OFDMA}}
\textcolor{NewColor}{
Fig.~\ref{fig:LWA-OFDM_vs_LWA-OFDMA} evaluates the differences in the sum rate obtained by using \ac{lwa} under \ref{itm:OFDM} and under \ref{itm:OFDMA}. 
For \ac{ofdm} where each subband can be used by all users \ref{itm:OFDM}, we note that the additional coding complexity is translated into improved sum rate, supporting the linear dependence of the rate in the term $\|\myVec{h}_{n}(b,L,\mySet{S}_n)\|^2$ according to \eqref{eq:1freq-rate}-\eqref{eq:sum rate}.
This is more evident as the number of users $K$ grows, as under \ref{itm:OFDMA}, each subband is designated to, at most, a single user. Still, \ac{ofdma} as in \ref{itm:OFDMA} is expected to be preferable in practice due to its notably reduced coding and processing complexity.  
In summary, Fig.~\ref{fig:LWA-OFDM_vs_LWA-OFDMA} validates that while \ac{ofdm}'s performance is ideal since its formulation considers multi-user coding, our proposed algorithm~\ref{alg:LWA opt} is  beneficial, up to some extent, also for \ac{ofdma}.
}

\begin{figure} % 4-8-16 users
		\centering
		\includegraphics[width=1\linewidth]{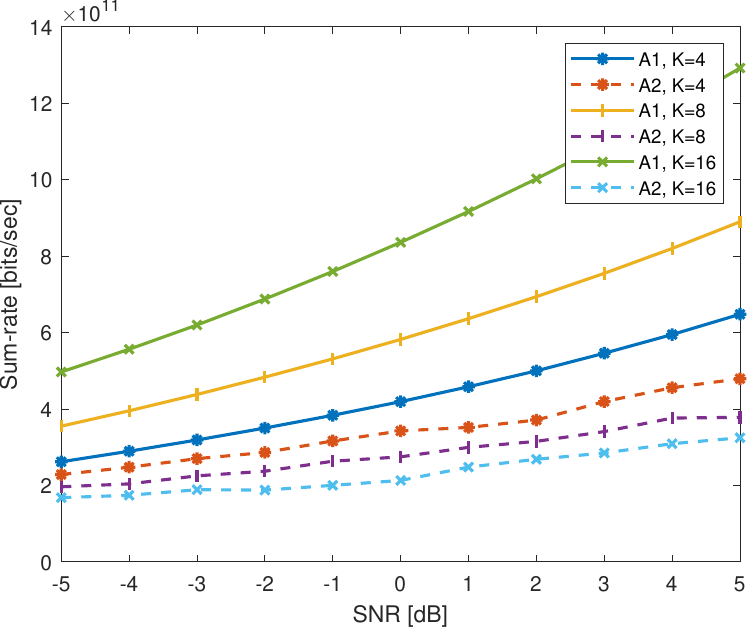}
		\caption{sum rate performance versus the \ac{snr} for $K\in\{4,8,16\}$ users, comparing \ac{lwa}-based \ac{ofdm} \ref{itm:OFDM} and \ac{ofdma} \ref{itm:OFDMA} signaling.} 
        \label{fig:LWA-OFDM_vs_LWA-OFDMA}
\end{figure}

\subsubsection{Comparison to MIMO Architectures}
\color{NewColor}
We proceed to comparing \acp{lwa} to the sum rate theoretically achievable with alternative \ac{mimo} architectures. As this study focuses on comparing different antennas, in these study we only include \ac{ofdm} for all architectures. Doing so allows us to capitalize on the different balances between cost and performance induced by the exploring different antenna architecture.   
\color{black}

%%%%%%%%% MIMO despciption
In particular, we compare the \ac{lwa} with two forms of \ac{mimo} architectures:
$(i)$ a fully digital wideband multi-antenna \ac{bs} equipped with a uniform linear array with $M$ half-wavelength spaced elements centered at the origin, corresponding to the antenna position in the \ac{lwa}-based channel; and
$(ii)$ a hybrid analog/digiral \ac{mimo} \ac{bs}, whose uniform linear array is fed using single RF chain (as in \acp{lwa}) using a standard phased-array interface, optimized using the method of \cite{ioushua2019family}.
It is emphasized that while phase-shifter based  hybrid analog/digital beamformers are widely considered in the literature as a low complexity architecture for THz communications~\cite{nguyen2024joint}, their associated complexity reduction is not in the same order as that provided by \acp{lwa}. For instance, phase shifters maintaining a constant phase and a unit amplitude across a wide frequency band (as that of THz communications) involves highly power consuming active circuitry needed to maintain the unit amplitude for all considered frequencies.

\textcolor{NewColor}{For the \ac{mimo} channels, the sum rate is given by the conventional expression for multi-band \ac{mimo} rate~\cite{brandenburg1974capacity}, obtained from the sum-capacity expression for broadcast channels with no shared messages (see \cite[Ch. 9.5]{el2011network}. The maximal achievable sum rate is then computed using standard spatial-spectral waterfilling~\cite{goldsmith2005wireless}.}
Since the channel model is tightly related to the antenna architecture, for the \ac{mimo} setting, we obtain the channel using the \ac{los} model of \cite{zhang2022beam} encompassing both radiative near- and far-field users. To guarantee a fair comparison, we normalize the \ac{mimo} channels to have the same max tap magnitude as the \ac{lwa} channel, and compute the average sum rate versus the \ac{snr}. We only normalize the \ac{mimo} channels to ensure that the optimization algorithm suggested in Subsection \ref{ssec:optimization} offers a solution independent of the channel normalization.

%Although our study in Subsection~\ref{subsec:SimBand} indicates that our design does not lead to a setting that violates the assumption of approximately flat sub-channels, we ensure that the rates we report hold, regardless of the maximal coherence width $d$ we choose. Specifically, we adjust the rate calculation to be based on the optimized power allocation of each individual algorithm, while resampling the proposed power allocation and the corresponding frequencies in the same thinly-divided frequency choice of the \ac{ofdm} setting, i.e., $f_n=f_{\rm min}+\frac{BW}{N}n,\ \forall n$. Calculating the sum rate this way assures that the resulting curves reported indeed hold (in the same sense as the computation of the \ac{ofdm} rate is) regardless of the maximal spectral width constraint that is imposed. 

\begin{figure*}
		\centering
        \begin{subfigure}{0.48\textwidth} % 4 users
         \includegraphics[width=0.9\linewidth]{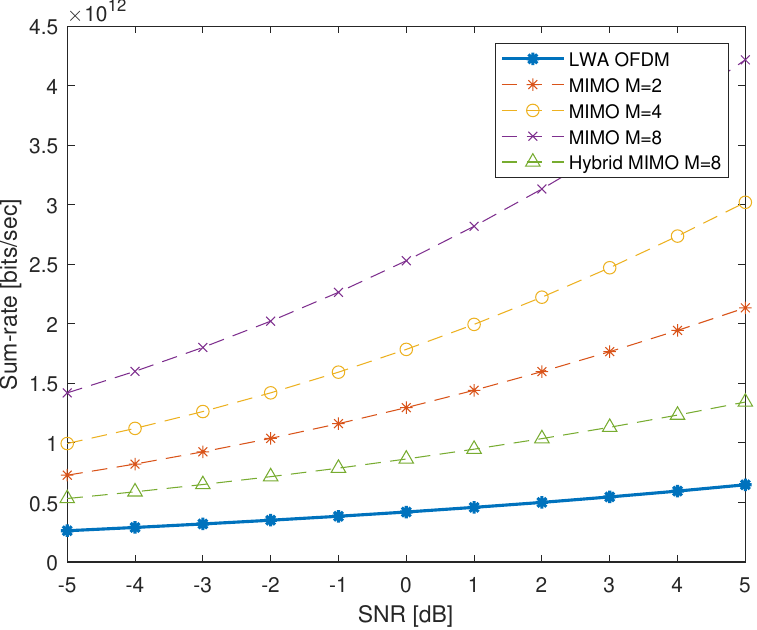} % good_beams OFDMA
         \caption{Small-scale multi-user scenario with $K=4$ users.}
         \label{fig:LWA_vs_MIMO_small-scale}  
         \end{subfigure}
         \hfill
         \begin{subfigure}{0.48\textwidth}  % 16 users
         \includegraphics[width=0.9\linewidth]{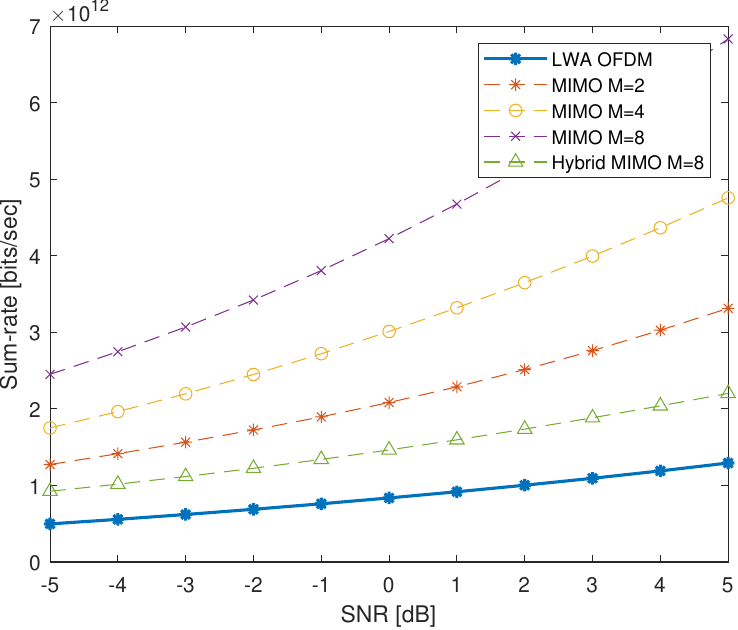} % bad_beams OFDMA
         \caption{Large-scale multi-user scenario with $K=16$ users.}
         \label{fig:LWA_vs_MIMO_large-scale}  
     \end{subfigure}
     \vspace{5mm}
    \caption{sum rate  versus  \acs{snr}, comparing \ac{lwa}-based \ac{ofdm} with fully digital and hybrid \ac{mimo} with $M$ transmit antennas.} 
    \label{fig:LWA_vs_MIMO}
\end{figure*} 

{\bf Small-Scale Multi-User Scenario.} 
The results reported in Fig.~\ref{fig:LWA_vs_MIMO_small-scale} show that for $K=4$, the sum rate achieved by a \ac{lwa}-based channel is comparable to the sum rates achieved by the \ac{mimo} setting, and may offer a valuable alternative to the (costly and currently non-feasible at THz) traditional setting, while supporting wideband transmissions. We note that in the \ac{mimo} setting, the beamforming is induced by manipulating the interference of correlated signals transmitted from multiple antennas, and thus the increase of the number of antennas, $M$, enables the forming of better directed beams and therefore increase the achievable rate. As a trade-off, the increase of $M$ is accompanied by an increase in the number of costly phase shifters which also consume additional power. 
 
{\bf Large-Scale Multi-User Scenario.}
We also repeat the sum rate comparison  for the large-scale scenario of $K=16$ users. The results depicted in Fig. \ref{fig:LWA_vs_MIMO_large-scale} show that the trends in terms of performance, compared to costly \ac{mimo} antennas noted for few users, hold when the number of users grows. Once again, we see that the \ac{lwa}-based channel supports communication rates that are comparable to the ones predicted under the \ac{mimo} setting. 

%Unlike the small-scale case, in the large-scale scenario, the \ac{lwa}-based channel demonstrates a different response to changes of the \ac{snr} compared to the \ac{mimo} channel. Specifically, the results associated with the \ac{lwa} show a faster increase in sum rate when increasing the \ac{snr}. 
%\textcolor{NewColor}{The change in response to \ac{snr} can be attributed to the varying angular distribution of users relative to the beams' widths. When users are sparsely distributed, each user is typically allocated different frequencies, corresponding to different beams. However, as the number of users grows, more users fall within the angular  range covered by a single frequency beam. This overlap may increase or decrease the effect of the \ac{snr}-based power allocation on the resulting rates.}

%----------------------------------------------------------------------------------------
%	CONCLUSIONS
%----------------------------------------------------------------------------------------
% \FloatBarrier
\section{Conclusions}
\label{sec:conclusions} 
In this paper, we studied wideband downlink multi-user communications at THz frequencies using the \ac{lwa} antenna technology. We introduced a model for \ac{lwa}-aided communications that faithfully captures its physical operations, while  encapsulating the ability of \acp{lwa} to generate frequency-selective directed THz beams with a single antenna element. We proposed an alternating optimization method for jointly tuning the \ac{lwa} configuration alongside the spectral power and frequency allocation. Our extensive numerical results showcased that a single \ac{lwa} can generate directed beams towards multiple users, and achieve sum rate performance comparable to a (extremely costly) fully digital \ac{mimo} array; for the conventional \ac{ofdm} and \ac{ofdma} settings. These results held for a small-scale scenario of $K=4$ users, as well as a large-scale one with $K=16$ users. 
We further demonstrate that our subband division, geared for \ac{ofdma}, achieves performance only slightly below that obtained when all users share the entire spectrum—a scenario that demands  advanced coding schemes.

The concept of leveraging the THz rainbow induced by \acp{lwa} to perform downlink communications with multiple users has only recently emerged, and have not yet been studied extensively. In this paper, we focused on unveiling the potential of this approach in performing high rate communications, compared to traditional \ac{mimo}, and analyzed the achievable rate under an ideal scenario of users that are non-selective towards frequency,  which can communicate simultaneously and are set at given locations. Our promising results set the ground for extensive future research, e.g.,  in formatting the paradigm for user localization, access control protocol, and rate optimization for frequency-selective users.

% %----------------------------------------------------------------------------------------
% %	APPENDICES
% %----------------------------------------------------------------------------------------%
 
% \begin{appendix} 
%     %
%     \numberwithin{proposition}{subsection} 
%     \numberwithin{lemma}{subsection} 
%     \numberwithin{corollary}{subsection} 
%     \numberwithin{remark}{subsection} 
%     \numberwithin{equation}{subsection}	
%     %
%     %
%     %-----------------------------------
%     %	Proof of waterfilling lemma
%     %----------------------------------- 
%     \subsection{Proof of Lemma \ref{lem:Waterfilling}}
%     \label{app:Proof1}
%     \textcolor{red}{TODO YAELA}

% \end{appendix}
 
%----------------------------------------------------------------------------------------
%	REFERENCES
%----------------------------------------------------------------------------------------

% \newpage
%\pagebreak
\bibliographystyle{IEEEtran}
\bibliography{IEEEabrv,mybib}

\end{document}